\documentclass[smus]{snow2e}
\usepackage{graphics}
\input epsf
\def\D0{D\O}
\def\etmisv {\mbox{${\hbox{${\vec E}$\kern-0.45em\lower-.1ex\hbox{/}}}_T$}}
\def\etmis  {\mbox{${\hbox{$E$\kern-0.45em\lower-.1ex\hbox{/}}}_T$ }}
\def\pbarp  {\mbox{$\overline{p}p$} }
\def\pdf    {parton distribution function }
\def\pdfs   {parton distribution functions }
\def\ifmath#1{\relax\ifmmode #1\else $#1$\fi}%
\def\GeV{\ifmmode {\mathrm{ Ge\kern -0.1em V}}\else
                   \textrm{Ge\kern -0.1em V}\fi}%
\def\MeV{\ifmmode {\mathrm{ Me\kern -0.1em V}}\else
                   \textrm{Me\kern -0.1em V}\fi}%
\def\keV{\ifmmode {\mathrm{ ke\kern -0.1em V}}\else
                   \textrm{ke\kern -0.1em V}\fi}%
\def\eV{\ifmmode  {\mathrm{ e\kern -0.1em V}}\else
                   \textrm{e\kern -0.1em V}\fi}%
\def\GeVcc{\ifmmode {\mathrm{ \GeV/c^2}}\else
                   \textrm{Ge\kern -0.1em V/c$^2$}\fi}%

\def\pbarp              {\mbox{$\overline{p}p$ }}
\def\ppbar              {\mbox{$\overline{p}p$ }}
\def\ttbar              {\mbox{$t\overline{t}$ }}

\newcommand{\AFB}       {A_{\mathrm{FB}}}
\newcommand{\ee}        {\mbox{$e^+e^-$}}
\newcommand{\qq}        {\mbox{$q\overline q$}}
\newcommand{\uu}        {\mbox{$u\overline u$}}
\newcommand{\dd}        {\mbox{$d\overline d$}}

\newcommand{\SM}        {\mbox{SM}}
\begin{document}

\begin{titlepage}
\rightline{\vbox{\halign{&#\hfil\cr
&Fermilab-Conf-96/353\cr
&November 1996\cr}}}
\vspace{1in}

\renewcommand{\thefootnote}{\fnsymbol{footnote}}
\setcounter{footnote}{2}

\begin{center}

{\Large\bf
    Electroweak Measurements from the
    Tevatron\footnote{\normalsize
    Invited talk given at the
    {\it Workshop on New Directions for High Energy Physics
    (Snowmass 96)},
    Snowmass, Colorado, June 25 -- July 12, 1996.}
}

\vskip 2.0cm
\normalsize
{\large Marcel Demarteau}
\vskip .3cm
Fermilab \\
Batavia, IL 60510, USA\\
\vskip 3cm

\end{center}

\begin{abstract}

An overview of recent electroweak physics results from the Tevatron 
is given. Properties of the $W^\pm$ and $Z^0$ gauge bosons
using final states containing electrons and muons based on large 
integrated luminosities are presented. In particular,
measurements of the $W^\pm$ and $Z^0$ production cross sections, the 
$W$-charge asymmetry and the measurement of the $W$-mass 
are summarized. Gauge boson self interactions are measured 
by studying gauge boson pair production and limits on anomalous gauge 
boson couplings are discussed. 

\end{abstract}

\setcounter{footnote}{0}
\renewcommand{\thefootnote}{\arabic{footnote}} \end{titlepage}

\pagestyle{plain}

\title{Electroweak Physics Results from the Tevatron
       \thanks{Work supported by the U.S. Dept. of Energy under 
               contract DEAC02-76CHO3000 }}

\author{ Marcel Demarteau \\
       {\it Fermi National Accelerator Laboratory} \\ 
       {\it P.O. Box 500, Batavia, IL 60565 }}

\maketitle


\begin{abstract}
An overview of recent electroweak physics results from the Tevatron 
is given. Properties of the $W^\pm$ and $Z^0$ gauge bosons
using final states containing electrons and muons based on large 
integrated luminosities are presented. In particular,
measurements of the $W^\pm$ and $Z^0$ production cross sections, the 
$W$-charge asymmetry and the measurement of the $W$-mass 
are summarized. Gauge boson self interactions are measured 
by studying gauge boson pair production and limits on anomalous gauge 
boson couplings are discussed. 
\end{abstract}

\section{Introduction }

The Standard Model of electroweak interactions
(SM) has taken a very prominent position in today's description of 
experimental results. 
Perhaps the most compelling reason for this state of affairs is that 
the experimental results have reached a level of precision which require 
a comparison with theory beyond the Born calculations, which the $\SM$ 
is able to provide. 
It is widely anticipated, though, that the $\SM$ is just an approximate 
theory and should eventually be replaced by a more complete and 
fundamental description of the underlying forces in nature. 
Since the highest center of mass energies are reached at the
Tevatron, the measurements at this accelerator provide natural tools 
to probe the $\SM$ at the highest energy scale. 

In this summary the most recent electroweak results 
from the Tevatron will be described, with the emphasis on 
results from the collider experiments CDF and \D0. 
The CDF and \D0 detectors are large multi-purpose detectors operating at 
the Fermilab Tevatron \pbarp Collider \cite{d0_nim,cdf_nim}. 
The \D0 detector has a non-magnetic inner 
tracking system, compact, hermetic, uranium liquid-argon calorimetry and 
an extensive muon system.
The CDF detector has a magnetic central detector, scintillator based 
calorimetry and a central muon system. 
During the 1992-1993 run, generally called Run~1a, 
the CDF and \D0 experiments have collected 
$\sim$20~pb$^{-1}$ and $\sim$15~pb$^{-1}$ of data, respectively. For the 
1994-1995 run (Run~1b) both experiments have collected 
$\sim$90~pb$^{-1}$ of data. 
The CCFR experiment at Fermilab studies $\nu_\mu$-nucleon interactions. 
The measurement of the ratio of charged and neutral current cross 
sections provides a direct measurement of the weak mixing angle. 
Results on the $W$ and $Z$ production cross sections, the $W$-width, 
$W$-charge asymmetry and the mass of the $W$-boson are presented. 
In the last section moments of the gauge boson 
are discussed.

\section{IVB Production Cross Sections }

In \pbarp collisions intermediate vector bosons are produced predominantly 
by quark-antiquark annihilation. In approximately 80\% of the interactions 
a valence quark is involved. Sea-sea interactions contribute
$\approx$20\% to the total cross section. The leptonic decay modes of the 
$W$ and $Z$-bosons are easily detected because of their characteristic decay 
signatures: for a $W$ decay a high $p_T$ lepton accompanied by 
large missing transverse energy 
(${\hbox{$E$\kern-0.45em\lower-.1ex\hbox{/}}}_T$), 
indicating the presence of a neutrino, and two high 
$p_T$ leptons for $Z$-decays. 
The measurement of the $W$ and $Z$ production cross sections probes the 
$\SM$ of electroweak and strong interactions and provides insight 
in the structure of the proton. With the large increase in integrated 
luminosity the new measurements have a significantly improved precision. 
A persistent uncertainty on any cross section measurement at a \pbarp 
collider, however, is the large uncertainty on the integrated luminosity 
due to the uncertainty on the effective total \pbarp cross section 
seen by the detectors. 
This uncertainty cancels completely in the ratio of the $W$ and $Z$ 
production cross sections, a quantity that 
can be used to extract the width 
of the $W$-boson, $\Gamma(W)$. 
The event selection is thus geared towards 
maximizing the cancellation of the different uncertainties in the ratio 
of the two cross section measurements. 

\begin{table}[h]
\begin{center}
{\footnotesize 
\begin{tabular}{||l|c|c||c|c||} \hline\hline
                & \multicolumn{2}{c||}{ \D0  }
                & \multicolumn{2}{c||}{ CDF  }  \\ \hline
                & $e$ & $\mu$      
                & $e$ & $\mu$   \\ \hline
$W$ cand.          &  59579         & 4472 
                   &  13796         & 6222           \\
A$_W$ (\%)         & 43.4 $\pm$ 1.5 & 20.1 $\pm$ 0.7 
                   & 34.2 $\pm$ 0.8 & 16.3 $\pm$ 0.4 \\
$\epsilon_W$ (\%)  & 70.0 $\pm$ 1.2 & 24.7 $\pm$ 1.5 
                   & 72.0 $\pm$ 1.1 & 74.2 $\pm$ 2.7 \\
Bkg (\%)           &  8.1 $\pm$ 0.9 & 18.6 $\pm$ 2.0 
                   & 14.1 $\pm$ 1.3 & 15.1 $\pm$ 2.2 \\
$\int {\cal L}$ (pb$^{-1}$) 
                   & 75.9 $\pm$ 6.4 & 32.0 $\pm$ 2.7 
                   & 19.7 $\pm$ 0.7 & 18.0 $\pm$ 0.7 \\ \hline 
$Z$ cand.          &  5702          & 173 
                   &  1312          & 423            \\
A$_Z$ (\%)         & 34.2 $\pm$ 0.5 &  5.7 $\pm$ 0.5 
                   & 40.9 $\pm$ 0.5 & 15.9 $\pm$ 0.3 \\
$\epsilon_Z$ (\%)  & 75.9 $\pm$ 1.2 & 43.2 $\pm$ 3.0 
                   & 69.6 $\pm$ 1.7 & 74.7 $\pm$ 2.7 \\
Bkg (\%)           &  4.8 $\pm$ 0.5 &  8.0 $\pm$ 2.1 
                   &  1.6 $\pm$ 0.7 &  0.4 $\pm$ 0.2 \\
$\int {\cal L}$ (pb$^{-1}$) 
                   & 89.1 $\pm$ 7.5 & 32.0 $\pm$ 2.7 
                   & 19.7 $\pm$ 0.7 & 18.0 $\pm$ 0.7 \\

&&&&\\    \hline\hline
\end{tabular}
}
\end{center}
\caption[]{Analysis results for the $W$ and $Z$-production cross section 
           measurement for CDF and \D0. 
           A$_V$, $\epsilon_V$ and Bkg stand for acceptance, detection 
           efficiency and Bkg, respectively, for vector boson $V$. }
\label{table:xsec}
\end{table}

$W$ and $Z$ events are normally recorded using a common single 
lepton trigger. 
The event selection for $W$-bosons requires an isolated lepton with 
transverse momentum $p_T > 25 $ GeV and $\etmis > 25 $ GeV. Leptonic 
decays of $Z$-bosons are selected by imposing the same lepton quality cuts 
on one lepton, and looser requirements on the second lepton. 
Table~\ref{table:xsec} lists the kinematic and geometric acceptance
(A$_V$), trigger and event selection efficiency ($\epsilon_V$) and 
background (Bkg) for the electron and 
muon decay channel for the two experiments ($V=W$ or $Z$) 
\cite{cdf_r,cdf_rmu,d0_xsec}.

\begin{figure}[t]
    \epsfxsize = 9.0cm
    \centerline{\epsffile{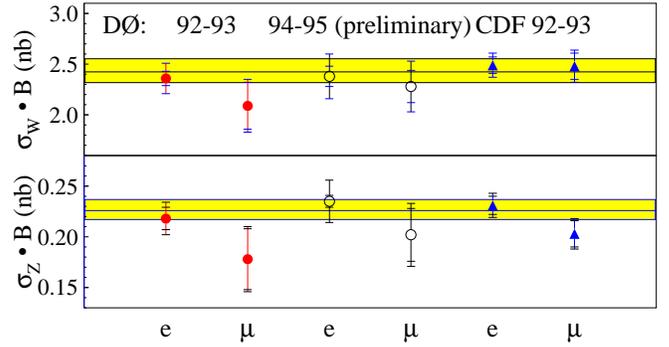}}
\caption{Measurements of the $W$ and $Z$ inclusive cross section compared 
         with the theoretical prediction using the CTEQ2M parton 
         distribution function. The shaded bands indicate the uncertainty 
         on the predictions. }
\label{fig:xsec}
\end{figure}

The vector boson inclusive cross section times decay branching ratio 
follows from the number of background subtracted observed candidate 
events, corrected for efficiency, acceptance and luminosity: 
\begin{displaymath} 
    \sigma \cdot B \,=\,  { N_{obs} - N_{bkg}   \over 
                            {\rm A} \, \epsilon \, {\cal L} } \ .
\end{displaymath} 
Here $N_{obs}$ is the observed number of events and $N_{bkg}$ the 
number of expected background events. $B$ indicates the 
branching ratio of the vector boson for the decay channel under study. 
The measured cross sections times branching ratio are listed in 
Table~\ref{table:xsec-result} and are compared with the theoretical 
prediction in Fig.~\ref{fig:xsec}. 
The theoretical predictions
for the total production cross section, 
calculated to ${\cal O}(\alpha_s^2)$ \cite{Neerven}, 
depend on three input parameters: 
the mass of the $W$-boson, taken to be 
$M_W = 80.23 \pm 0.18$ GeV/c$^2$, the mass of the $Z$-boson, 
$M_Z = 91.188 \pm 0.002$ GeV/c$^2$~\cite{old_masses}, 
and the structure of the proton. Using the CTEQ2M parton distribution 
functions~\cite{cteq}, 
the prediction for the total cross sections are 
$\sigma_W$ = 22.35~nb and $\sigma_Z$ = 6.708~nb. 
Using the leptonic branching ratio 
$B(W \rightarrow \ell\nu) = (10.84 \pm 0.02)\% $, 
as calculated following reference \cite{Rosner_w_width} using the above 
quoted $W$-mass, and 
$B(Z \rightarrow \ell\ell) = (3.366 \pm 0.006)\%$ as measured by the LEP 
experiments \cite{pdg}, 
the theoretical predictions for the total inclusive production 
cross section times branching ratio are 
$\sigma_W \cdot B(W \rightarrow \ell\nu)  = 2.42^{+0.13}_{-0.11}$~nb and 
$\sigma_W \cdot B(Z \rightarrow \ell\ell) = 0.226^{+0.011}_{-0.009}$~nb. 
The two largest uncertainties on the theoretical prediction are the 
choice of \pdf (4.5\%) and the uncertainty due to using a NLO \pdf with a 
full ${\cal O}(\alpha_s^2)$ theoretical calculation (3\%). The experimental
error is dominated by the uncertainty on the luminosity.

\begin{table}[h]
\begin{center}
{\footnotesize 
\begin{tabular}{||l|l|l||} \hline\hline
      & \multicolumn{1}{c|}{ 
        $\sigma_W \cdot B (W\rightarrow \ell\nu)$ }
      & \multicolumn{1}{c||}{ 
        $\sigma_Z \cdot B (Z\rightarrow \ell\ell)$ }	   \\ \hline
1992-1993          &                                   &   \\
\D0 (e)            & 2.36  $\pm$ 0.02  $\pm$ 0.15 
                   & 0.218 $\pm$ 0.008 $\pm$ 0.014         \\
\D0 ($\mu$)        & 2.09  $\pm$ 0.06  $\pm$ 0.25 
                   & 0.178 $\pm$ 0.022 $\pm$ 0.023         \\
CDF (e)            & 2.49  $\pm$ 0.02  $\pm$ 0.12
                   & 0.231 $\pm$ 0.006 $\pm$ 0.011         \\
CDF ($\mu$)        & 2.48  $\pm$ 0.03  $\pm$ 0.16
                   & 0.203 $\pm$ 0.010 $\pm$ 0.012         \\ \hline 
1994-1995          &                                   &   \\
\D0 (e)            & 2.38  $\pm$ 0.01  $\pm$ 0.22 
                   & 0.235 $\pm$ 0.003 $\pm$ 0.021         \\
\D0 ($\mu$)        & 2.28  $\pm$ 0.04  $\pm$ 0.25 
                   & 0.202 $\pm$ 0.016 $\pm$ 0.026         \\
&&\\    \hline\hline
\end{tabular}
}
\end{center}
\caption[]{Measured cross section times branching ratio in nb for 
           $W$ and $Z$ production based on integrated luminosities of 
           12.8 (11.4) pb$^{-1}$ and 19.7 (18.0) pb$^{-1}$ 
           for the electron (muon) channel for \D0 and CDF, respectively 
           for the 1992-1993 data and the preliminary \D0 results 
           for 75.9 (32.0) pb$^{-1}$ of data from the 1994-1995 run. }
\label{table:xsec-result}
\end{table}

\begin{figure}[h]
    \epsfxsize = 7.0cm
    \centerline{\epsffile{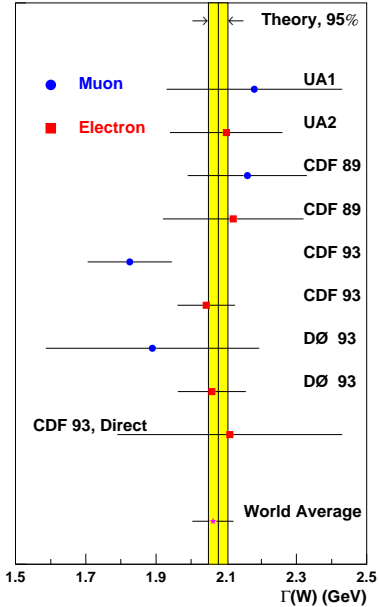}}
\caption{Measurements of $\Gamma_W$ compared with the $\SM$ 
         expectation. }
\label{fig:gamma_w}
\end{figure}

The ratio of the cross section measurements in which the error on the 
luminosity, common to both the $W$ and $Z$ events, completely cancels 
measures the leptonic branching ratio of the $W$-boson. 
It can be used, within the above framework, to extract
the total width of the $W$-boson: 

\begin{displaymath} 
    R \,=\,  { \sigma_W \cdot B(W\rightarrow\ell\nu)  \over
               \sigma_Z \cdot B(Z\rightarrow\ell\ell) } 
      \,=\,  { \sigma_W \over \sigma_Z } \cdot 
             {  \Gamma(W\rightarrow\ell\nu )  \over 
                \Gamma(Z\rightarrow\ell\ell ) } 
             {  \Gamma(Z) \over \Gamma(W) } 
\end{displaymath} 
which gives 
\begin{displaymath} 
    B^{-1}(W\rightarrow\ell\nu)  \,=\, 
            { \sigma_W \over \sigma_Z }                 \cdot 
            { 1        \over B(Z\rightarrow\ell\ell) }  \cdot 
            { 1        \over R } 
\end{displaymath} 
Using the $\SM$ prediction \cite{Rosner_w_width} for the partial 
decay width 
$\Gamma(W\rightarrow\ell\nu)$ the total width $\Gamma_W$ is given by  
\begin{displaymath} 
    \Gamma_W \,=\, 
                { \sigma_W \over \sigma_Z }     \cdot 
                { \Gamma(W\rightarrow\ell\nu)   \over 
                  B(Z\rightarrow\ell\ell) }     \cdot 
                { 1 \over R } 
\end{displaymath} 
The ratio of the cross sections, using again the calculation of 
\cite{Neerven}, is determined to be 3.33 $\pm$ 0.03. The error is again 
dominated by the choice of parton distribution functions. 
Note that in the ratio the 
theoretical uncertainties also largely cancel. 
Using, as before, the measured branching ratio 
$B(Z \rightarrow \ell\ell) = (3.366 \pm 0.006)\%$ and the 
theoretical prediction for the partial decay width 
$\Gamma(W\rightarrow\ell\nu)$ = 225.2 $\pm$ 1.5 MeV 
\cite{Rosner_w_width}  
the $W$ leptonic branching ratio, as determined from the 
combined \D0 electron and muon 
1992-1993 data, is (11.02 $\pm$ 0.5)\%; the CDF measured branching
ratio, based on the 1992-1993 electron data is 
(10.94 $\pm$ 0.33 $\pm$ 0.31)\%. Using the calculated 
partial leptonic branching ratio, 
these measurements yield for the width 
$\Gamma_W~=~2.044~\pm~0.093$ GeV  \cite{d0_xsec} and 
$\Gamma_W~=~2.043~\pm~0.082$ GeV \cite{cdf_r}, respectively. 
The CDF value differs from their published value due to the use of more 
recent experimental measurements in evaluating the input parameters. 
Figure~\ref{fig:gamma_w} shows the world $W$-width measurements together 
with the theoretical prediction 
\cite{cdf_r,d0_xsec,ua1_gamma_w,ua2_gamma_w}.

Taking into account that the ratio of the total cross sections
$\sigma_W /\sigma_Z$ is slightly different at a center of mass energy of 
630~GeV 
($\sigma_W /\sigma_Z (\sqrt{s}=630$ GeV) = 3.26 $\pm$ 0.09), 
and accounting for the 
correlation between the measurements at different center of mass energies 
through the choice of parton distribution functions, the different values 
of $\Gamma_W$ can be combined to give a world average of 
$\Gamma_W = 2.062 \pm 0.059$ GeV, a measurement at the 3\% level. 
This is in good agreement with the $\SM$ prediction of 
$\Gamma(W) = 2.077 \pm 0.014$ GeV. The comparison of the measurement with 
the theoretical prediction can be used to set an upper limit 
on an \lq\lq excess width\rq\rq\ 
$\Delta \Gamma_W \equiv \Gamma_W{\rm (meas)} - \Gamma_W {\rm  (SM)}$,
allowed by experiment for  non--$\SM$ decay processes, such 
as decays into supersymmetric particles or into heavy quarks. 
Comparing the above world average value of $\Gamma_W$ with the 
$\SM$ prediction a 95\% C.L. upper limit of 
$\Delta\Gamma < 109$~MeV on unexpected decays can be set.

Since the intermediate vector bosons are produced through a Breit--Wigner 
resonance the line shape of the mass distribution contains information 
about the width of the boson. 
For $W$-bosons, the high tail of the 
transverse mass distribution, where the Breit-Wigner shape dominates 
over the detector resolutions, can be used to extract $\Gamma_W$. 
Using a binned log--likelihood method, CDF has fit the transverse 
mass\footnote{Transverse mass is defined as the invariant mass of the lepton 
              and the neutrino of the $W$-decay in the transverse plane 
              (see section \ref{sec_wmass}). } 
($m_T$) distribution 
far above the $W$ pole ($m_T > 110$~GeV/c$^2$) 
to Monte Carlo generated templates with varying $W$-width 
\cite{cdf_gamma_w_direct}. 
Using this 
method the $W$-width has been determined to be 
$\Gamma_W = 2.11 \pm 0.28 \pm 0.16$~GeV, 
where the systematic error (8\%) is dominated by uncertainties in 
modelling the $W$ transverse momentum distribution (6\%)  and the 
\etmis resolution (5\%). Although the precision of this method is currently 
not competitive with the extraction of the width from the ratio of 
cross sections, it has the advantage that it is relatively independent 
of $\SM$ assumptions.

\section{Drell-Yan Production }

One of the unique features of \pbarp collisions is the large range of 
available partonic center of mass energies. This allows for a study 
of the $Z$ line shape through the Drell-Yan process
($\qq \rightarrow (\gamma, Z \rightarrow ) \ \ell^+\ell^- )$ over 
a large di-lepton invariant mass region.
The low invariant mass region 
allows access to the small $x$ region of the parton distribution 
functions down to $x=0.006$, where $x$ is the fraction of the proton 
momentum carried by the parton. 
The region well above the $Z$ pole is the region where 
the $\gamma Z$ interference effects are strongest. 
A possible substructure of the partons
would manifest itself most prominently in a modification of the
interference pattern. Substructure of partons is most commonly
parametrized in terms of a contact interaction~\cite{contact},
characterized by a phase, $\eta$, leading to constructive ($\eta=-1$) 
or destructive interference ($\eta=+1$) with the $\SM$ Lagrangian, and a 
compositeness scale, $\Lambda_\eta$, indicative of the 
energy scale at which substructure would be revealed. 
By fitting the di-lepton invariant mass spectrum 
to various assumptions for the compositeness 
scale and the phase of the interference, 
lower limits on the compositeness scale can be set.

\begin{figure}[t]
    \epsfxsize = 8.0cm
    \centerline{\epsffile{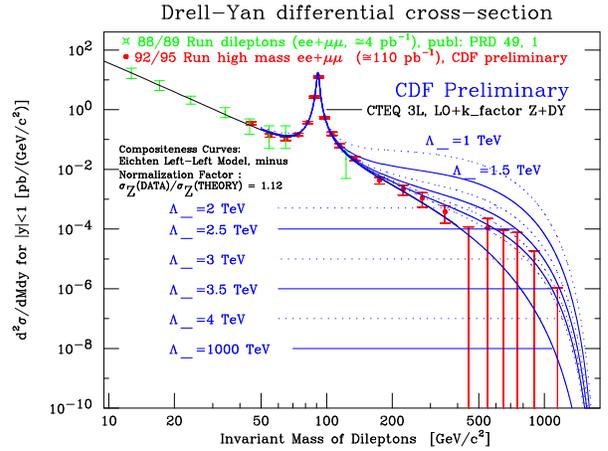}}
\caption{Double differential cross section $d^2\sigma / dM\,dy$ 
         for CDF electron and muon data combined. The open symbols 
         are from the 88/89 data. The solid symbols correspond to the 
         full Run I data. The curves are the theoretical predictions 
         for different $\Lambda_-$ values. }
\label{fig:cdf_dy}
\end{figure}

The CDF experiment has measured the double differential Drell-Yan
cross section $d^2\sigma / dM\,dy$ for electron and muon pairs
in the mass range $11 < M_{\ell\ell} < 150$ GeV/c$^2$ for the
Run~1a data~\cite{cdf_dy1a}, and
$40 < M_{\ell\ell} < 550$ GeV/c$^2$ for the Run~1b data. 
The di-electron invariant mass spectrum is measured over the rapidity 
interval $|\eta|<1$. Due to a more restricted coverage, the muon 
cross section has been determined only over the range $|\eta|<0.6\,$.
Figure~\ref{fig:cdf_dy} shows the measured cross section for electrons
and muons combined. 
The curves correspond to a leading-order calculation of the
Drell-Yan cross section with in addition a contact interaction of
left-handed quarks and leptons with positive interference for 
different values of the compositeness scale. 
Higher order effects have been included through the use 
of a constant $k$-factor of $k=1.12$\, . 
The curve for $\Lambda_-=1000$~TeV indicates the $\SM$ prediction. 
The data is clearly inconsistent with low $\Lambda_-$ values. 
Performing a maximum likelihood fit yields scale 
factors for the electron data of 
$\Lambda_-$    $\geq$     3.4 TeV,    
$\Lambda_+$    $\geq$     2.4 TeV   and for the muon data of
$\Lambda_-$    $\geq$     3.5 TeV, 
$\Lambda_+$    $\geq$     2.9 TeV. Combining both channels yields 
$\Lambda_+$    $\geq$     2.9 TeV   and 
$\Lambda_-$    $\geq$     3.8 TeV. 
This implies that up to a distance of $< 10^{-17}$ cm the 
interacting particles reveal no substructure.

\section{Forward-Backward Asymmetry }

Because the left-handed and right-handed coupling of fermions to the 
$Z$ boson are not the same, the angular distribution of the 
outgoing fermion with respect to the incoming fermion in the 
parton center of mass frame, has 
a term linear in $\cos\vartheta^*$~\cite{angle}. 
The angular distribution is thus 
asymmetric and will exhibit a forward-backward asymmetry, defined as 
\begin{displaymath} 
    \AFB \, = \, {\sigma_F \,-\, \sigma_B 
                  \over 
                  \sigma_F \,+\, \sigma_B }
\end{displaymath} 
where $\sigma_F$ is the cross section for fermion production in the
forward hemisphere ($0^\circ < \vartheta^* < 90^\circ$) and, 
correspondingly, $\sigma_B$ for the backward hemisphere 
($90^\circ < \vartheta^* < 180^\circ$).
Due to the changing polarization of the $Z$ boson as function of center 
of mass energy, $\AFB$ has a strong energy dependence. Since 
the couplings of the fermions to the $Z$ boson depend on the fermion 
weak isospin and charge, $\AFB$ is different for different 
initial and final states. For the Drell-Yan process 
$\ppbar \rightarrow \ell^+\ell^- $ no distinction can be made between 
$\uu$ and $\dd$ initial states and therefore the asymmetry 
measured will be a convolution of both. It is interesting to note 
that this process is the time-reversal of the corresponding process 
at $e^+e^-$-machines and the measurements are complementary. 
At LEP and SLC the measurements are free from \pdf uncertainties, whereas 
at the Tevatron, the light quark asymmetries are free from fragmentation 
uncertainties. 

\begin{figure}[h]
    \epsfxsize = 7.0cm
    \centerline{\epsffile[0 200 600 700]{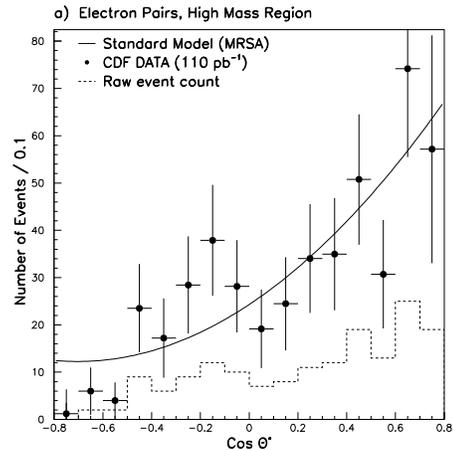}}
    \hspace*{-1.0cm}
\caption{Distributions in $\cos\vartheta^*$ for events from the process 
         $\pbarp \rightarrow Z/\gamma + X$, $Z/\gamma \rightarrow \ee$ 
         for the di-electron invariant mass region 
         $M_{ee} > $ 105~GeV/c$^2$. 
         The points are the fully corrected data and the line is the
         $\SM$ calculation, normalized to the number of events 
         observed in the data. The dashed histogram is the raw
         event count. }
\label{fig:cdf_afb}
\end{figure}

The CDF experiment has measured $\AFB$ using the full Run~I data set
for di-electron final states with 
$|\eta_{\ell_1}| < 1.1 $ and $|\eta_{\ell_2}| < 2.4$~\cite{cdf_afb}. 
The data sample is divided into two invariant mass regions: a 
pole region, 75 $< M_{ee} < $ 105 GeV/c$^2$ with 5463 events and a 
high mass sample, $M_{ee} > $ 105 GeV/c$^2$ with 183 events. 
Figure~\ref{fig:cdf_afb} shows the event count in $\cos\vartheta^*$ 
for the high mass sample. The dashed line is the raw data distribution 
and already shows a clear forward-backward asymmetry. 
The points are the corrected data
compared to the $\SM$ prediction using the MRSA 
parton distribution function~\cite{mrs}. 
The background in the pole-region is dominated by QCD di-jet events 
where both jets either contain or fake an electron. It has been 
estimated to be $110 \pm 36$~events. 
In the high mass region the background is relatively small 
but has a large uncertainty, $0^{+21}_{-0}$~events, which dominates 
the systematic uncertainty on the measurement in this mass region. 
Because of the finite mass resolution, events will migrate between the two 
mass regions. The deconvolution of the mass resolution 
is performed with a Monte Carlo
simulation and results in a correction on $\AFB$ of 
$\Delta_{\AFB} = +0.07  \pm 0.03$ in the high mass region and 
$\Delta_{\AFB} = -0.010 \pm 0.003$ in the pole region. The
corrections for angular acceptance have also been determined from 
Monte Carlo simulations. 
The analysis yields 
$A_{FB} = 0.07 \pm 0.016 $ for 
75 $< M_{ee} < $ 105 GeV/c$^2$, and 
$A_{FB} = 0.43 \pm 0.10 $ 
for $M_{ee} > $ 105 GeV/c$^2$, compared to the $\SM$ predictions
of $A_{FB} = 0.054 \pm 0.001 $ and 
$A_{FB} = 0.528 \pm 0.006 $, respectively. 

\begin{figure}[h]
    \epsfxsize = 7.0cm
    \centerline{\epsffile{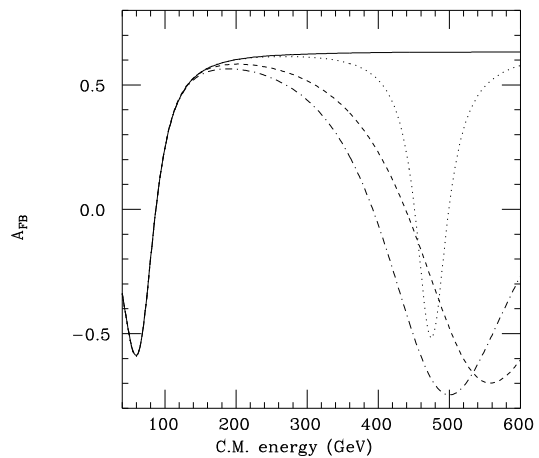}}
\caption[]{Parton level forward-backward asymmetry as function of center of 
           mass energy for $\dd \rightarrow e^+e^-$ for the
           $\SM$ (solid line), and for models with an additional 
           $Z_I$ (dashed-dotted line), $Z_\chi$ (dashed line) or 
           $Z_\psi$ (dotted line) boson of 
           500~GeV/c$^2$~\cite{rosner_zprime}.}
\label{fig:zprime_afb}
\end{figure}

Even though in the high mass region the asymmetry is measured with a
rather large error, these measurements still serve as a probe of 
extensions of the $\SM$ because models with additional heavy
neutral gauge bosons can substantially alter $\AFB$. 
For example, Fig.~\ref{fig:zprime_afb} from~\cite{rosner_zprime} 
shows $\AFB$ for $\dd \rightarrow e^+e^-$ as function of the 
partonic center of mass energy for the $\SM$ (solid line) 
and for various models with an additional neutral heavy gauge boson
with a mass of 500~GeV/c$^2$. 
A modest event sample at a center of mass energy of 
${\sqrt {\hat s}} = M_{Z^\prime}$ allowing an unambiguous sign 
determination of $\AFB$, would already put constraints on extended 
gauge sectors in the $\SM$.

\section{$W$-mass }
\label{sec_wmass}

The mass of the $W$-boson is one of the fundamental parameters of the 
$\SM$. A precision measurement of the $W$-boson mass allows 
for a stringent test of the 
radiative corrections in the $\SM$. 
Combined with the measurement of the mass of the top-quark and precision 
measurements from $e^+e^-$ and neutrino scattering experiments,
inconsistencies between the different measurements 
can be looked for, possibly indicating processes beyond the 
$\SM$. 

In $W$ events produced in a hadron collider in essence only two 
quantities are measured: the
lepton momentum and the transverse momentum of the recoil system. 
The latter consists of the ``hard'' $W$-recoil and the underlying event
contribution. For $W$-events these two are inseparable. 
The transverse momentum 
of the neutrino is then inferred from these two observables. Since the
longitudinal momentum of the neutrino cannot be determined unambiguously, 
the $W$-boson mass is determined from the line shape in 
transverse mass, 
defined as 
\begin{displaymath} 
    m_T \,=\,  \sqrt{ 2\, p_T^\ell \, p_T^\nu \, 
                     (1 - \cos\varphi^{\ell\nu}) } \ .
\end{displaymath} 
Here $\varphi^{\ell\nu}$ is the angle between the lepton and neutrino in 
the transverse plane. 

Both the transverse mass and lepton transverse momentum are, by
construction, invariant under longitudinal Lorentz boosts. 
The quantity transverse mass is preferred over the lepton transverse 
momentum spectrum because to first order it is independent of the 
transverse momentum of the $W$. Under transverse Lorentz boosts 
along a direction $\varphi^*$, 
$m_T$ and $p_T^\ell$ transform as
\begin{eqnarray*} 
    M_T^2 &\cong & {M_T^*}^2 \,-\, \beta^2 \, \cos^2\varphi^* \, {M_L^*}^2
    \\
    p_T^\ell &\cong & {p_T^\ell}^* \,+\, \frac{1}{2} \, 
                                         \beta \, \cos\varphi^*  \, M_W 
\end{eqnarray*} 
with 
$M_T^* = M_W \, \sin\vartheta^* $, 
$M_L^* = M_W \, \cos\vartheta^* $  and 
$\beta = {p_T^W \over M_W} $. 
The asterisks indicate quantities in the $W$ rest frame. 
The lepton transverse momentum depends linearly on $\beta$ whereas the
dependence of the transverse mass is second order in $\beta$. 
The disadvantage of using the transverse mass is that it uses the 
neutrino transverse momentum which is a derived quantity. The neutrino 
transverse momentum is equated to the missing transverse energy in the 
event, which is given by 
\begin{eqnarray*}
    \etmisv \,=\, - {\displaystyle \sum_i } \, {\vec p}_{T_i}
            \,=\, -   {\vec p}_T^{\,e} \,-\, {\vec p}_T^{\,rec} 
                \,-\, {\vec u}_T({\cal L})
\end{eqnarray*}
where ${\vec p}_T^{\,rec}$ is the transverse momentum of the $W$-recoil 
and ${\vec u}_T({\cal L})$ the transverse energy flow of the 
underlying event, which depends on the luminosity. 
It then follows that the magnitude of the missing 
$E_T$ vector and the true neutrino momentum are related as 
$ \etmis \,=\, p_T^\nu \,+\, {1\over 4} \, {u^2_T \over p_T^\nu} $. 
This relation can be interpreted as the definition of the neutrino
momentum scale. Note that the underlying event gives rise to a bias 
in the measured 
neutrino momentum with respect to the true neutrino momentum. 
When there are more interactions per crossing 
$| {\vec u}_T |$ behaves as a two-dimensional 
random walk and is proportional to $\sqrt{I_C}$, where 
$I_C$ is the number of interactions per crossing. The shift in measured 
neutrino momentum is thus directly proportional to the number of 
interactions per crossing. The resolution increases as $\sqrt{I_C}$. 
At high luminosities alternate methods to determine the $W$-mass 
may therefore be advantageous~\cite{mw_other}. 

Since there is no analytic description of the transverse mass distribution, 
the $W$-mass is determined by fitting Monte Carlo generated templates 
in transverse mass for different masses of the $W$-boson to the data 
distribution. This distribution exhibits a Jacobian edge characteristic
of two-body decays which contains most of the mass information. 
For the $W$-mass determination both the 
energy scale for the lepton and recoil system, which determine the 
peak position of the transverse mass distribution, as well as 
the resolutions on the measured variables, which control the
steepness of the Jacobian edge, are crucial. 

The CDF mass analysis discussed here is based on the Run~1a 
data~\cite{cdf_mw}. 
The \D0 mass analysis also includes a preliminary result from the Run~1b
data~\cite{d0_mw}. 
In the CDF $W$-mass analysis 
the momentum scale of the central
magnetic tracker is set by scaling the measured $J/\psi$-mass to 
the world average value using $J/\psi \rightarrow \mu^+\mu^-$ decays. 
Based on a sample of approximately 60,000 events a scale factor of 
0.99984~$\pm$~0.00052 has been derived. 
The dominant contribution to the 
error comes from the uncertainty in the amount of material the muons
traverse. This procedure establishes the momentum scale at the
$J/\psi$-mass, where the average muon $p_T$ is about 3~GeV/c, and needs to
be extrapolated to the momentum range appropriate for leptons from 
$W$-decays. The error due to possible 
nonlinearities in the momentum scale is addressed by studying the 
measured $J/\psi$-mass as function of $\langle 1/p_T^2 \rangle$, 
extrapolated to zero curvature. 
This extrapolation, which includes an uncertainty on a possible 
non-linearity of the momentum measurement, increases the error 
on the momentum scale to 0.00058 at the $W$-mass. 
This results in an error on the $W$-mass 
of 50~MeV/c$^2$. 

\begin{figure}[t]
    \epsfxsize = 9.0cm
    \centerline{\epsffile[50 400 550 700]{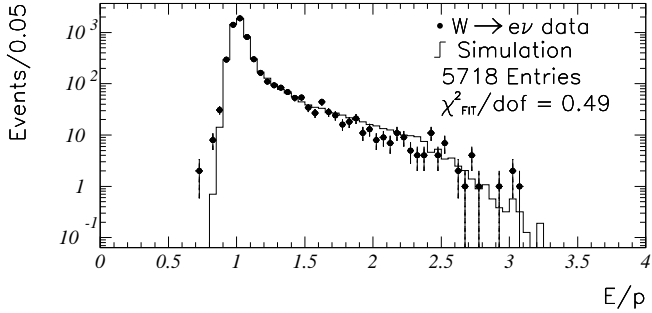}}
\caption{The $E/p$ distribution for electrons in the $W$-sample (points) with 
         the best fit from the simulation (histogram). }
\label{fig:cdf_ep} 
\end{figure}

Having established the momentum scale, the calorimeter energy scale 
is determined from a line shape comparison of the observed $E/p$
distribution with a detailed Monte Carlo prediction as shown in 
Fig.~\ref{fig:cdf_ep}. A two-dimensional fit 
of Monte Carlo generated $E/p$ distributions in the energy scale and the 
electron momentum resolution is used to establish the absolute calorimeter 
energy scale. The scale factor is extracted from a fit over the range 
$0.9~<~E/p~<~1.1$. Since the momentum measurement is very sensitive to
bremsstrahlung effects, the energy scale determination is critically
dependent on an accurate modelling of the amount of material the electrons
traverse. Using the ratio of events in the region $1.3 < E/p < 2.0$ to the 
events in the range $0.8 < E/p < 1.2$ the amount of material is determined 
to be (8.9$\pm$0.9)\% $X_0$, consistent with independent checks using photon 
conversions and $Z$-events but slightly higher than from a direct 
accounting of the material. The limited statistics in the high $E/p$
region is the dominant source of the systematic error on the amount of 
material traversed by electrons and thus on the energy scale
determination. The uncertainty of 10\% on the amount of material in front 
of the calorimeter contributes a 70 MeV/c$^2$ uncertainty on the $W$-mass. 
The other two main contributions to the total energy scale error are a 
65~MeV/c$^2$ error due to the statistics in the $E/p$-peak and a 
50~MeV/c$^2$ error from the uncertainty on the electron resolution. 
The total error on the $W$-mass from setting the energy scale using 
the momentum scale is thus 110~MeV/c$^2$ which, 
combined with the 50~MeV/c$^2$ momentum scale uncertainty, gives a total 
energy scale uncertainty on the $W$-mass of 120~MeV/c$^2$ for the
measurement using $W\rightarrow e\nu$ decays. 

The energy and momentum scales are verified by measuring the masses of
known resonances, the $Z$-mass and the masses of the $\Upsilon$
resonances. They are all in good agreement with the world average values. 
The width of the $Z$-resonance provides a constraint on the momentum
resolution that results in a systematic error on the $W$-mass from the
uncertainty on the momentum and energy resolution of 60~MeV/c$^2$ and 
80~MeV/c$^2$ for the muon and electron measurement, respectively. 
The hadronic energy scale does not need to be determined separately since 
$Z\rightarrow e^+e^-$ collider events are used to model the $W$-recoil
system. 

At \D0 the $W$-mass is measured from $W\rightarrow e\nu$ decays. 
The electromagnetic (EM) energy scale is determined by calibrating to 
the $Z\rightarrow ee$ resonance. Since the absolute energy scale 
of the EM calorimeter is not known with the required precision,
the ratio of the measured $W$ and $Z$ masses and the world average
$Z$ mass are used to determine the $W$ boson mass.
The $W$ mass measured is de facto the ratio of the measured 
$W$ and $Z$ mass, scaled to the LEP $Z$ mass: 
$M_W = {M_W^{\D0} \over M_Z^{\D0}} \times M_Z^{\rm LEP}$. 
A number of systematic effects, common to both measurements, cancel
in the ratio. Most notably, as shall be discussed in more detail below, 
the ratio is to first order insensitive to the absolute energy scale. 

Test beam measurements have demonstrated the EM calorimeter to be 
linear to better than 0.5\% for electron energies exceeding 10~GeV. 
To establish the energy scale with the precision required for this 
measurement, it is necessary to determine to which extent a potential 
offset in the energy response, as opposed to a scale factor, is
responsible for the deviation of the ratio 
$M_Z^{\D0} \over M_Z^{\rm LEP}$ from unity. 
This was achieved by combining the measured $Z$ mass with the
measurements of 
$\pi^0 \rightarrow \gamma\gamma$ and $J/\psi \rightarrow e^+e^-$
decays and comparing them to their known
values~\cite{ref_mass}.
If the electron energy measured in the calorimeter and its true energy
are related by
$E_{\rm meas} = \alpha \, E_{\rm true} + \delta$,
the measured and true mass values are, to first order, related
by $m_{\rm meas} = \alpha \, m_{\rm true} \,+\, \delta \, f $.
The variable $f$ depends on the decay topology and is given by
$f = {2(E_1 + E_2) \over m_{\rm meas} } \sin^2\gamma/2$,
where $\gamma$ is the opening angle between the two decay products
and $E_1$ and $E_2$ are their measured energies.

\begin{figure}[h]
    \epsfxsize = 8.0cm
    \centerline{\epsffile[0 150 600 600]{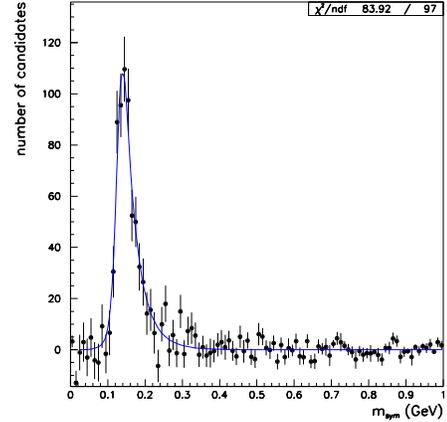}}
\caption[]{The $M_{sym}$ mass spectrum obtained from 
           $\pi^{0} \rightarrow \gamma\gamma$ decays. }
\label{fig:pizero}
\end{figure}

\begin{figure}[h]
    \epsfxsize = 7.0cm
    \centerline{\epsffile[0 200 600 600]{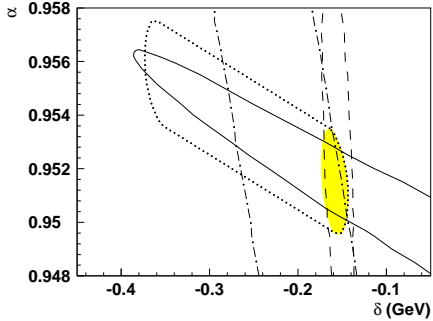}}
\caption[]{Constraints on slope $\alpha$ and intercept $\delta$ from
           observed $J/\psi \rightarrow e^+e^-$ (dashed-dotted line),
           $\pi^0 \rightarrow \gamma\gamma$ (dashed line),
           and $Z \rightarrow e^+e^-$ decays (solid line).
           The shaded inner contour shows the combined result.
           The dotted line indicates the allowed area when
           nonlinear terms, as constrained by test beam measurements,
           are included. }
\label{fig:contour}
\end{figure}

Figure~\ref{fig:pizero} shows the background subtracted mass spectrum 
of the decay $\pi^0 \rightarrow \gamma\gamma$. 
The two photons in the decay of the neutral pion are not resolved in the
calorimeter, but by selecting events in which both $\gamma$'s convert 
into an $e^{+}e^{-}$-pair, 
and produce distinctive doubly ionizing tracks in the central
detector, the opening angle can be reconstructed.
The ``mass" plotted in Fig.~\ref{fig:pizero}
(data points with error bars) is
\begin{equation}
    M_{sym} =  E \cdot \sin\frac{\vartheta}{2} ,
\end{equation}
where $E$ is the cluster energy, equal to the sum of the photon energies,
and $\vartheta$ is the opening angle of the two photons. 
$M_{sym}$ is equal to the invariant mass for symmetric decays.
The shape compares well with the Monte Carlo simulation shown as the
solid line. The measured mass is $M_{\pi^0} = (135.4 \pm 10.0)$~MeV/c$^2$.
The sensitivity to the energy scale and offset is determined by varying
both parameters in a Monte Carlo simulation and performing a 
$\chi^2$ fit to the data. 
This procedure maps out an allowed region in the 
($\alpha,\delta$)-plane shown as the dashed line in 
Fig.~\ref{fig:contour}. 

Similarly, a $J/\psi$ signal with a 
significance of about 5$\sigma$ has been extracted from the data, 
which yields an additional, independent constraint on $\alpha$ and
$\delta$ (dashed-dotted line in Fig.~\ref{fig:contour}). 
The strongest constraint on the energy scale uncertainty comes from 
the $Z$ data. The fact that electrons from $Z$ decays are not
monochromatic is exploited by studying the invariant mass distribution as
function of the variable $f$. Small values of $f$ correspond to the decay
of highly boosted $Z$ bosons with, on average, higher energies. The
dependence of the observed $Z$ boson mass as function of $f$ thus
directly translates into a constraint on the energy scale and offset, 
shown as the solid line in Fig.~\ref{fig:contour}. Each of the mass
states has a different sensitivity to $\alpha$ and $\delta$ and, taken 
together, provide a powerful tool for establishing the energy scale in
situ. 
When combined, these three constraints
limit $\alpha$ and $\delta$ to the shaded elliptical region.
Test beam measurements permit a small nonlinear term in the
energy response, which affects both $\alpha$ and $\delta$
and alters the ratio $M_W / M_Z$ largely through the effect on
$\delta$. The allowed region in the ($\alpha,\delta$)-plane 
when nonlinearities are included is 
indicated by the dotted line in Fig.~\ref{fig:contour}.
                                                                         
Using the measured masses for the observed resonances, the energy
scale factor determined for the Run Ia data is $\alpha =
0.9514 \pm 0.0018 {}^{+0.0061}_{-0.0017}$ and the offset is $\delta =
(-0.158 \pm 0.015 {}^{+0.03}_{-0.21})$~GeV, where the asymmetric errors
are due to possible calorimeter nonlinearities.
The measured offset is consistent with that determined from test beam
data,
and has been confirmed by a detailed Monte Carlo study of
energy loss in the central detectors.
The dependence of the measured ratio of the $W$ mass to $Z$ mass on
$\alpha$ and $\delta$ may be estimated from
\begin{eqnarray*}
    \left. \frac{M_W (\alpha,\delta)}{M_Z (\alpha,\delta)}\right|_{\rm meas}
           =
    \left. \frac{M_W}{M_Z}\right|_{\rm true}
           \left[ 1 + \frac{\delta}{\alpha} \cdot
                      \frac{f_W \, M_Z - f_Z \, M_W}{M_Z \cdot M_W}
                      \right] \ . 
\end{eqnarray*}
It should be noted that the $W$ mass is insensitive to $\alpha$ if
$\delta=0$.
The offset results in a $5~\rm{ MeV/c^2}$ correction to the measured 
$W$ mass. The uncertainty on the absolute energy scale
results, for the Run~Ia data sample, in an uncertainty on $M_W$ of 
160~MeV/c$^2$, of which 150~MeV/c$^2$ is due to the limited $Z$
statistics. For the Run Ib data sample, with a total integrated
luminosity of approximately 76~pb$^{-1}$, the energy scale uncertainty 
on the $W$ mass is 80~MeV/c$^2$. 
                                                                         

The $W$ event sample is selected by placing very stringent kinematic and
fiducial cuts. 
Both the CDF and \D0 mass analyses are currently based on event samples with 
central leptons only. The main difference in event selection is the 
treatment of the hadronic activity in the event. The CDF event selection 
excludes events with jets with $E_T > 30$~GeV. In addition $p_T^W$ 
is required to be less than 20~GeV/c, whereas \D0 only requires 
$p_T^W < 30$~GeV/c. 
These sets of selection criteria 
yield event samples of 8049 and 4663 events for the electron 
and muon decays, respectively, for CDF and 
7234 $W\rightarrow e\nu$ decays for the Ia and 
32856 for the Ib data set for \D0. 

The $W$-mass is then determined from a maximum likelihood fit 
of Monte Carlo generated templates in transverse mass to the 
data distributions. 
In the Monte Carlo model of $W$-production, events are generated according 
to a relativistic Breit-Wigner resonance, with a longitudinal momentum
distribution as given by the chosen parton distribution function. 
The CDF choice for nominal \pdf is the MRSD$^\prime$- pdf~\cite{mrs}. 
In their model the transverse momentum of 
the $W$ is generated according to the measured 
$p_T$ distribution of $Z$-events. This procedure can be justified because 
of the similarity between $W$ and $Z$-production and because there are
large uncertainties, both theoretical as well as experimental, on the 
$W$ $p_T$-distribution. 
The procedure has an added advantage that 
the recoil system does not need to be modeled independently, since it 
is taken directly from $Z$-events with the two leptons removed. 
This recoil distribution from $Z$-events is corrected for the lepton 
removal and modified to match data and Monte Carlo with respect to 
the width of the distribution of the 
projection of the $p_T$ of the recoil system perpendicular to the 
lepton direction. 
The disadvantage of the method is that very few events (555 events to 
be precise) are
used to model the recoil with a slightly different acceptance than for 
$W$-events, and it ignores the correlation between the
transverse and longitudinal momenta and the difference in mass between 
the $W$ and $Z$-bosons.

\begin{figure}[t]
    \epsfxsize = 7.0cm
    \centerline{\epsffile{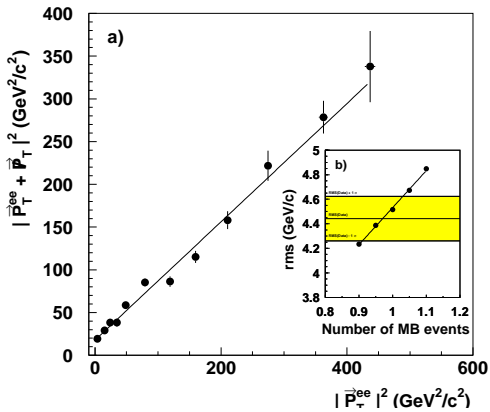}}
\caption[]{a) Distribution of
              $ | {\vec p}_T^{\,e_1} + {\vec p}_T^{\,e_2} + \etmis |^2 $
              versus $ | {\vec p}_T^{\,ee} |^2 $ for $Z$ events;
           b) Sensitivity of the width of the distribution in 
              ${\vec p}_T^{\,ee} + \vec{p}_T^{\,rec} + {\vec u_T}$, 
              projected 
              along the bi-sector of the two electrons,
              on the number of minimum bias events. The band corresponds 
              to the $\pm\,1\sigma$ uncertainty on this measurement. }
\label{fig:urecoil}
\end{figure}

The \D0 experiment generates $W$ bosons using the double
differential production cross section in $p_T$ and rapidity
calculated at next to leading order~\cite{ly} using the MRSA 
parton distribution functions~\cite{mrs}, thus including the 
correlation between the longitudinal and transverse momentum. 
Minimum bias events are used to model the underlying event,
mimicking the debris in the event due to spectator parton
interactions and the pile-up associated with multiple interactions,
and including the residual energy from previous beam crossings.
The relative response of the hadronic and EM calorimeters
is established by studying $Z$ events.
To ensure an equivalent event topology between the $W$ and $Z$ events,
$Z$ decays in which one electron is in the end calorimeter are included
in this study.
The transverse momentum balance in $Z$ events is given by
$ {\vec p}_T^{\,e_1} + {\vec p}_T^{\,e_2} + {\vec p}_T^{\,rec}
                     + {\vec u_T}
     = - {\hbox{$\rlap{\kern0.20em/}\vec E_T$}} $. 
One finds for the average
$    | {\vec p}_T^{\,e_1} + {\vec p}_T^{\,e_2} +
   {\hbox{$\rlap{\kern0.20em/}\vec E_T$}} |^2 =
     \kappa^2 \, | {\vec p}_T^ {\,ee} |^2 + | {\vec u_T} |^2 $
assuming
$ | {\vec p}_{T}^{\,rec} | = \kappa \, | {\vec p}_{T}^{\,ee} | $,
where $ {\vec p}_{T}^{\,ee} $ is the transverse momentum of the $Z$
measured from the two electrons.
The cross term on the right hand side averages to zero since
the underlying event vector is randomly distributed with respect
to the $Z$ recoil system.
Figure~\ref{fig:urecoil}a shows the distribution of
$ | {\vec p}_T^{\,e_1} + {\vec p}_T^{\,e_2} +
{\hbox{$\rlap{\kern0.20em/}\vec E_T$}} |^2 $
versus $ | {\vec p}_T^{\,ee} |^2 $.
The data shows a linear relation between the EM and hadronic energy scale, 
and yields $\kappa = 0.83 \pm 0.04$.
The intercept yields the magnitude of the
underlying event vector, $|{\vec u_T}| = 4.3 \pm 0.3 ~\rm{GeV/c}$,
consistent with the value obtained from minimum bias events.
The uncertainty on $M_W$ due to the uncertainty on the hadronic energy
scale is 50~MeV/c$^2$ for the Run Ia data.

The modeling of the recoil and  underlying event
are verified and constrained
by comparing the $p_T$ of the $Z$ obtained from
the two electrons, $\vec{p}_T^{\,ee}$, to that
obtained from the rest of the event:  $-\vec{p}_T^{\,rec} - {\vec u_T}$.
To minimize the contribution from the electron energy resolution,
the vector sum of these two quantities is projected
along the bisector of the two electron directions.
Since $\vec u_T$ is randomly oriented and has a magnitude
$\sim p_T^Z$, the width of the distribution is sensitive
to the underlying event contribution while the mean is largely unaffected.
The sensitivity of the width of this distribution to the mean number
of minimum bias events that mimic the underlying event is determined
by varying the number of minimum bias events in the Monte Carlo, 
as shown by the points in Fig.~\ref{fig:urecoil}b.
For the Ia data, the number of minimum bias events preferred is 
$0.98 \pm 0.06$, consistent with one.
The uncertainty on $M_W$ from the underlying event model is 60~MeV/c$^2$.

\begin{figure}[t]
\begin{center}
\begin{tabular}{c}
    \epsfxsize = 7.0cm
    \epsffile{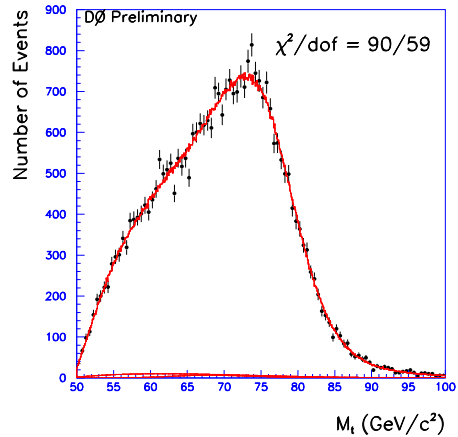}
\end{tabular}
\end{center}
\caption{\D0 transverse mass distribution of $W\rightarrow e\nu$ decays 
         collected during the 1994-1995 run. The points are the data 
         and the line is the best fit. } 
\label{fig:mt_d0} 
\end{figure}

\begin{figure}[t]
\begin{center}
\begin{tabular}{c}
    \epsfxsize = 8.0cm
    \epsffile{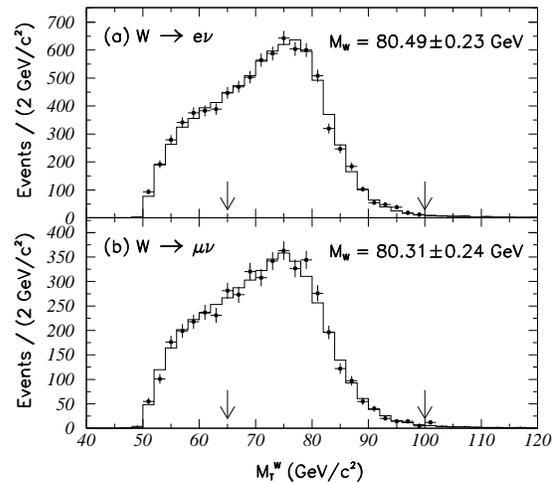}   
\end{tabular}
\end{center}
\caption{Transverse mass distribution of $W\rightarrow e\nu$ (top) and 
         $W\rightarrow \mu\nu$ (bottom) decays from CDF. The points are 
         the data and the histogram is the best fit to the data. The 
         arrows indicate the range used to extract the $W$-mass. }
\label{fig:mt_cdf} 
\end{figure}

The mass of the $W$ is obtained from a maximum likelihood fit over 
the transverse mass range 
$ 65 < m_T < 100 $ GeV/c$^2$ ($ 60 < m_T < 90 $ GeV/c$^2$)
for CDF (\D0). 
Figures~\ref{fig:mt_d0} and~\ref{fig:mt_cdf} 
show the transverse mass distributions for the data
together with the best fit of the Monte Carlo 
for the Run Ib electron data for \D0 and 
for the muon and electron channel for Run~1a for CDF, respectively. 
The $W$-mass is determined to be 
$M_W^\mu = 80.310 \pm 0.205 (stat) \pm 0.130 (sys)$~GeV/c$^2$ based on 3268 
$W\rightarrow \mu\nu$ events in the mass fitting window and 
$M_W^e   = 80.490 \pm 0.145 (stat) \pm 0.175 (sys)$~GeV/c$^2$ based on 5718 
events for CDF. 
\D0 finds 
$M_W^e = 80.350 \pm 0.140 \ {\rm (stat.)} \pm 0.165 \ {\rm (syst.)}
                \pm 0.160 \ {\rm (scale)}$~GeV/c$^2$ based on 5982
events in the mass fitting window using the Ia data, and 
$M_W^e = 80.380 \pm 0.070 \ {\rm (stat.)} \pm 0.130 \ {\rm (syst.)}
                \pm 0.080 \ {\rm (scale)}$~GeV/c$^2$ based on 27040
events for the Ib data. 
Table~\ref{table:mw_sys} lists the systematic errors on the individual
measurements and the common errors. 

\begin{table*}[th]
\begin{center}
\begin{tabular}{||l|rrr|rrr||} \hline\hline
    &   \multicolumn{3}{c|}{ CDF }
    &   \multicolumn{3}{c||}{ \D0 }                             \\ \hline
    &   \multicolumn{1}{c}{ e }
    &   \multicolumn{1}{c}{ $\mu$ }
    &   \multicolumn{1}{c|}{ common }
    &   \multicolumn{1}{c}{ Ia }
    &   \multicolumn{1}{c}{ Ib }
    &   \multicolumn{1}{c||}{ common }
\\ \hline
Statistical              & 145   & 205   &  ---  & 140 &  70 & ---    \\
Energy scale             & 120   &  50   &   50  & 160 &  80 &  25    \\
Angle scale              & ---   & ---   &  ---  &  50 &  40 &  40    \\
$E$ or $p$ resolution    & 80    &  60   &  ---  &  70 &  25 &  10    \\
$p_T^W$ and recoil model & 80    &  75   &   65  & 110 &  95 &        \\
pdf's                    & 50    &  50   &   50  &  65 &  65 &  65    \\
QCD/QED corr's           & 30    &  30   &   30  &  20 &  20 &  20    \\
$W$-width                & 20    &  20   &   20  &  20 &  10 &  10    \\
Backgrounds              & 10    &  25   &  ---  &  35 &  15 & ---    \\
Efficiencies             &  0    &  25   &  ---  &  30 &  25 & ---    \\
Fitting procedure        & 10    &  10   &  ---  &   5 &   5 & ---  \\ \hline
Total                    & 230   & 240   &   100 & 270 & 170 &  80    \\
\hline\hline
Combined                 & \multicolumn{3}{c|}  { 180 }
                         & \multicolumn{3}{c||} { 150}                \\
\hline\hline
\end{tabular}
\end{center}
\caption[]{Errors on $M_W$ in MeV/c$^2$. }
\label{table:mw_sys}
\end{table*}

The dominant theoretical uncertainty in this measurement comes from the 
$p_T^W$ model and the uncertainty on the proton structure. 
Parton distributions and the spectrum in $p_T^W$ are correlated. 
The \D0 experiment has addressed this correlation in the determination
of its uncertainty on the $W$ mass. 
In their analysis new parametrizations of the
CTEQ~3M parton distribution function were obtained that included in the
fit the CDF $W$ asymmetry data from Run~Ia~\cite{cdf_wasym}, 
where all data points had
been moved coherently up or down by one standard deviation. 
In addition one of the parameters, which describes the
$Q^2$-dependence of the parametrization of the non-perturbative
functions describing the $p_T^W$ spectrum~\cite{ly}, was varied. 
The constraint on this parameter was provided by the measurement of 
the $p_T^Z$ spectrum. 
The uncertainty due to parton distribution functions and the 
$p_T^W$ input spectrum was then assessed by varying simultaneously 
these new parton distribution function 
and the parameter describing the 
non-perturbative part of the $p_T^W$ spectrum. 
A total error on the $W$-mass of 65~MeV/c$^2$ has been assigned due to
these uncertainties. 

\begin{figure}[t]
    \epsfxsize = 8.0cm
    \centerline{\epsffile{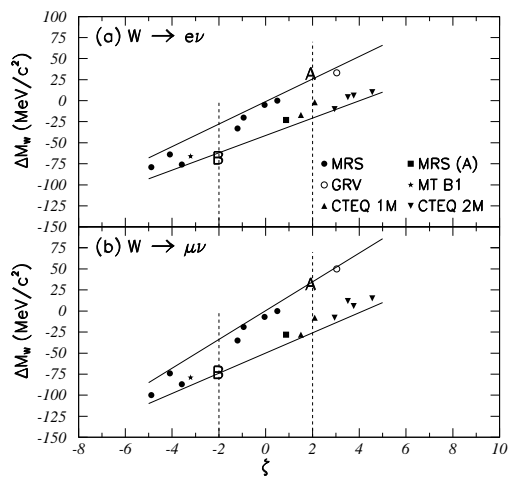}}
\caption{Correlation between $\Delta M_W$ and $\zeta$, the significance of 
         the difference between data and theory for the $W$-charge asymmetry, 
         for various \pdfs for the 
         (a) $W\rightarrow e\nu$- and (b) $W\rightarrow \mu\nu$-sample. 
         The nominal mass measurement uses the MRSD$^\prime-$ parton
         distribution function. }
\label{fig:cdf_mw_pdf} 
\end{figure}

The CDF experiment uses their measurement of the $W$ charge asymmetry 
as the sole constraint on the uncertainty due to the $p_T^W$ and 
parton distribution functions. 
Figure~\ref{fig:cdf_mw_pdf} shows the
correlation between $\Delta M_W$ and the significance of the deviation 
of the theoretical prediction for the $W$-asymmetry and the data 
for the electron and muon channel separately (cf.~eq.~(\ref{eq:sig_asym})). 
The uncertainty on $M_W$ is taken to be the 
symmetrized spread in masses for $-2 < \zeta < 2$, being 
50~MeV/c$^2$. 

Combining~\cite{mw_combined} 
these measurements with previous $W$ mass measurements~\cite{mw_recent},
assuming the only correlated uncertainty between the measurements is 
due to the parton distribution functions, gives a world average of 
$M_W = 80.356 \pm 0.125$~\GeVcc.

An indirect measurement of the $W$-mass, through the measurement of the 
weak mixing angle $\sin^2\vartheta_W$, is obtained from the study of
$\nu$N deep inelastic scattering experiments. 
The CCFR experiment studies $\nu_\mu$-nucleon interactions
and the ratio of charged and neutral current cross 
sections provides a direct measurement of the weak mixing angle. 
The cross sections have large contributions from electroweak 
radiative corrections. In the \lq\lq on shell\rq\rq\ scheme, however, 
where $\sin^2\vartheta_W$ is defined as $1-{M_W^2 \over M_Z^2}$ to 
all orders,
these corrections largely cancel in the ratio, thus reducing the 
dependence on the top mass and Higgs mass significantly and providing 
an indirect measurement of $M_W$. A preliminary value of 
$\sin^2\vartheta_W = 0.2213 \pm 0.0021 ({\rm stat.}) 
                            \pm 0.0027 ({\rm syst.}) 
                            \pm 0.0034 ({\rm model}) $
has been reported~\cite{ccfr_moriond}, corresponding to a $W$ mass 
value of $M_W = (80.46 \pm 0.25)$~GeV/c$^2$. The largest contribution 
to the systematic uncertainty comes from the uncertainty on 
the flux of background $\nu_e$'s. The model uncertainty is dominated 
by the turn-on of the charm quark production 
cross section. The latter uncertainty is expected to be reduced 
substantially with the follow-up experiment NuTeV, 
which will be able to measure the cross sections with 
neutrino and anti-neutrino beams separately.

\section{$W$-charge Asymmetry }

As Fig.~\ref{fig:cdf_mw_pdf} shows, the $W$ mass is strongly correlated
with the parton distribution functions. 
The parton distribution functions can be 
constrained at the appropriate $Q^2$-scale by measuring the charge 
asymmetry in $W$-production itself. The two, partly compensating, 
sources that contribute to the 
$W$-charge asymmetry are the production and decay processes. 
Since on average a $u$-quark carries more momentum than a 
$d$-quark, more $W^+$-bosons are produced  along the proton direction
than along the anti-proton direction resulting in a production charge 
asymmetry defined as 
\begin{eqnarray*} 
    A(y_W) \,&=&\,  { dN^+ (y_W) / dy \,-\, dN^- (y_W) / dy \over 
                      dN^+ (y_W) / dy \,+\, dN^- (y_W) / dy }  
\label{eq:asym}
\end{eqnarray*} 
The $W$-rapidity, $y_W$, however, 
cannot be reconstructed unambiguously because 
of the two-fold ambiguity in the longitudinal momentum of the neutrino. 
The quantity that is measured experimentally is the decay lepton charge 
asymmetry, defined as 
\begin{displaymath} 
    A(y_\ell) \,=\,  
              { dN^+ (y_\ell) / dy_\ell \,-\, dN^- (y_\ell) / dy_\ell \over 
                dN^+ (y_\ell) / dy_\ell \,+\, dN^- (y_\ell) / dy_\ell } 
\end{displaymath} 
where $N^{+(-)}$ is the number of positively (negatively) charged leptons 
detected at pseudorapidity $y_\ell$. 
Since the rapidity of the decay lepton is measured, there is an 
additional contribution from the $V-A$ coupling of 
the $W$. Since $W$-bosons are produced through 
$\qq$ annihilation they are almost fully polarized and 
the lepton from, for example, the $W^+$-decay is preferentially 
emitted along the anti-proton direction, 
which partially undoes the production asymmetry. 
Because of ${\cal CP}$ symmetry, $A(+y) \,=\, -\,A(-y)$, the measured 
asymmetries at positive and negative rapidities can be combined to get a
statistically more powerful measurement. 
The $V-A$ structure of the $W$-decay is very well understood. Thus, the 
charge asymmetry measurement can be used to probe the structure of the 
proton in the $x$ range 0.007 to 0.27$\,$. 

The CDF experiment, based on an integrated luminosity of about 20~pb$^{-1}$ 
measured the charge asymmetry for $W$-decays into electrons and muons 
and constrained the then current 
parton distribution functions \cite{cdf_wasym}. 
The lepton pseudorapidity range in that analysis was 
$|\eta| < 1.0$ for muons and $|\eta| < 2.4$ for electrons. 
It was limited by the rapidity coverage provided by the central
tracking chamber. 
The analysis has been updated~\cite{cdf_summer96} 
using the full Run~1 data set with a
total integrated luminosity of 110~pb$^{-1}$. 
The rapidity coverage for muons has been extended by utilizing the 
forward muon toroids~\cite{aha} 
covering $1.95 < |\eta| < 3.6$, which collected
72~pb$^{-1}$ of data. The efficiency for electrons in the
plug calorimeter ($1.1 < |\eta| < 2.4$) was also substantially improved.
In the previous analysis only the central tracking chamber was used in
the electron identification. 
Because of the limited coverage of this tracking system almost 
no tracks were reconstructed beyond $|\eta | \approx 1.8$\,. 
In the new analysis, utilizing the silicon vertex detector 
(SVX) and the vertex chamber, an average track finding
efficiency of 60\%, almost uniform in $\eta$, has been obtained out
to rapidities of $|\eta | \approx 2.3$\,. 
For the high $\eta$ region, though, the electron charge cannot be 
determined by the tracking system alone. 
In this region the charge is determined from a comparison of the
$\varphi$-angle as determined from the SVX track, and from the calorimeter
energy deposition. At the location of the calorimeter an average
displacement of 0.5~cm is expected in the pseudorapidity range 
$1.2 < |\eta| < 1.8$, which is measured with a 
resolution of 0.15~cm. 

\begin{figure}[t]
    \epsfxsize = 8.0cm
    \centerline{\epsffile{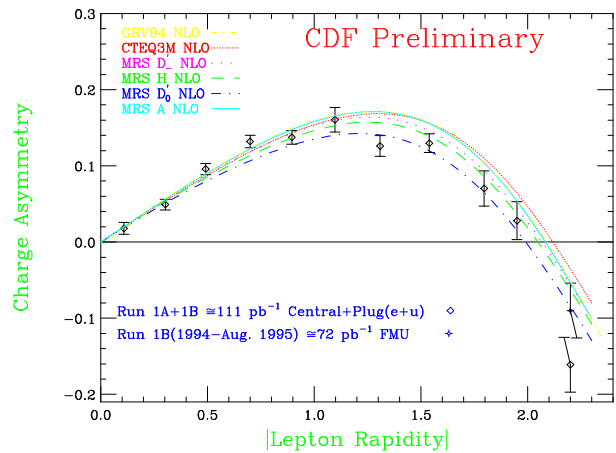}}    
\caption{CDF Run~I measured lepton charge asymmetry from 
         $W\rightarrow \ell\nu$ events compared to NLO 
         predictions for different parton distribution functions. }
\label{fig:cdf_wasym} 
\end{figure}

Figure~\ref{fig:cdf_wasym} shows the measured asymmetry 
as a function of the lepton rapidity together with
the theoretical prediction for different parton distribution functions. 
The predictions were obtained using the DYRAD NLO Monte Carlo
\cite{dyrad}.  Compared to the previous analysis the new measurements at 
high rapidity should be noted. 
Since the measurement is a ratio measurement, many systematic errors 
cancel and the total systematic error is about 20\% of the statistical 
error.

The asymmetry measurement provides an independent discriminant 
between different parton distribution functions. The disagreement
between theory and experiment can be quantified by defining the
significance of the disagreement between the weighted mean asymmetry 
($\overline A$) from theory and experiment as 
\begin{equation} 
    \zeta \,=\, { \overline{A}_{\,pdf} \,-\, \overline{A}_{\,data} \over 
                  \sigma(\overline{A}_{\,data})   }  \ \ .
\label{eq:sig_asym} 
\end{equation} 
The $\zeta$ values listed in Table~\ref{tab:wasym} seem to prefer 
the recent MRS \pdfs \cite{mrs} over other 
distributions~\cite{cteq,grv}. 
The constraint which the $W$ charge asymmetry
provides on the uncertainty on the $W$ mass measurement, however, is 
not expected to scale with event statistics, since the measurement is
mainly sensitive to the slope of the ratio of the $u$ and $d$ parton
distribution functions and does not probe the full parameter range
describing them. 

\begin{table}[h]
\begin{center}
\begin{tabular}{|l|c|} \hline
    PDF Set         & $\zeta$   \\
\hline
    CTEQ3M          & 1.16      \\
    MRS A, G        & 1.75      \\
    MRS H           & -0.51     \\
    MRSD$^\prime$-  & 0.68      \\
    GRV 94          & 2.59      \\
    GRV 92          & 4.13      \\
\hline
\end{tabular}
\caption[]{Comparison between measured and predicted asymmetry for 
           different parton distribution functions. } 
\label{tab:wasym}
\end{center}
\end{table}


\section{Rare $W$ Decays }

The study of rare decays provides a precision test of the
underlying theory since in general the predictions of rare decay 
rates involve higher order calculations. 
$W$ decays into a pseudoscalar meson and a photon, $W\to P\gamma$,
are particularly attractive since they are sensitive to new
physics which affects the $WW\gamma$ vertex. A search for $W\to P\gamma$
decays thus complements di-boson analyses described in detail in
the following section.  

\begin{figure}[t]
    \epsfxsize = 8.0cm
    \centerline{\epsffile[125 260 625 700]{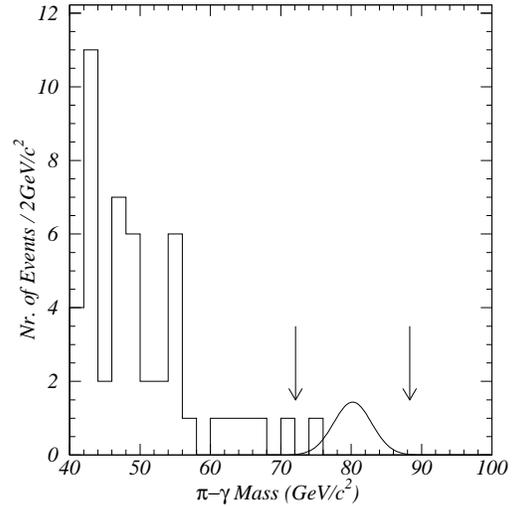}}
\caption{The CDF distribution in $M(\pi\gamma)$ for the search for the 
         rare decay $W\rightarrow \pi\gamma$. The arrows indicate the
         search window. The Gaussian, centered at $M_W$, corresponds to
         the 95\% CL limit of 4.9 events. } 
\label{fig:rarew} 
\end{figure}

Currently, experiments have only looked for the rare decay 
$W\to\pi\gamma$~\cite{cdf_rarew_89,ua2_rarew,cdf_rarew_96} with the
strongest limit coming from the latest CDF analysis. 
In this analysis, based on an integrated luminosity of 16.7~pb$^{-1}$,
events were selected with an energetic photon and a single central jet
with $E_T > 15$~GeV with a matching isolated track. The track was
required to have $p_T > 15$~GeV/c with no other charged tracks with
$p_T > 1$~GeV/c in a cone of radius $\Delta R = 0.7$\,.
By initially not placing a cut on the electromagnetic fraction of the 
pion jet, the sample is dominated by isolated electrons and permits 
measurement of many of the efficiencies from the data itself. 
In the final selection the electromagnetic
fraction of the jet is required to be less than 80\% of the total jet 
energy, and a sample of 79 events remains (see Fig.~\ref{fig:rarew}) 
with one event in
the search region $|M(\pi\gamma) - M_W | < 8.1$~\GeVcc. 

The background, primarily coming from jet production with the
jet opposite the photon candidate fragmenting into a single charged
particle, possibly associated with neutrals, has been estimated to be 
$2.6 \pm 1.0 \pm 1.3$~events in the mass window. The one event observed
is thus consistent with background. Without background subtraction, the
95\% confidence level limit is 4.9~events. Using the measured 
$W$ production cross section, this results in a 95\% CL upper limit 
on the partial decay width of 
\begin{displaymath} 
    {\Gamma ( W \rightarrow \pi^\pm \gamma ) 
     \over 
     \Gamma ( W \rightarrow e\nu ) } 
     \,<\, 2 \cdot 10^{-3} \ \ , 
\end{displaymath} 
to be compared with the theoretical prediction of~\cite{rarew}
$ \Gamma ( W \rightarrow \pi^\pm \gamma ) / 
  \Gamma ( W \rightarrow e\nu )  \sim 3 \cdot 10^{-8}$.

\section{Gauge Boson Pair Production }

Similar to a study of rare decays of vector bosons, a study of the 
magnetic dipole and electric quadrupole moments of the $W$ boson 
probes the $W$ interaction vertex. 
The non-Abelian $SU(2)\times U(1)$ gauge symmetry of the $\SM$ implies 
that the gauge bosons self-interact. These self-interactions give rise 
to very subtle interference effects in the $\SM$ and the couplings are 
uniquely determined by the gauge symmetry in
order to preserve unitarity. 
The magnetic dipole and electric quadrupole moments of the $W$ are, in 
the $\SM$ at tree level, given by: 
\begin{displaymath} 
    \mu_W = {e \over m_W} \qquad Q_W^e = {-e \over m_W^2} \ .
\end{displaymath} 

The most general effective electroweak Lagrangian, invariant under 
$U(1)_{\rm EM}$, however, contains eight independent 
coupling parameters, the ${\cal CP}$--conserving parameters $\kappa_V$ and 
$\lambda_V$ and the ${\cal CP}$--violating parameters 
$\tilde{\kappa}_V$ and $\tilde{\lambda}_V$, 
where $V$ = $\gamma$ or $Z$. 
The ${\cal CP}$--conserving parameters are related to the magnetic dipole 
($\mu_W$) and electric quadrupole ($Q_W^e$) moments of the $W$ boson, while
the ${\cal CP}$--violating parameters are related to the electric dipole
($d_W$) and the magnetic quadrupole ($Q_W^m$) moments~\cite{Kim}: 
\begin{eqnarray*} 
    \mu_W &=& (e/2m_W)(1+\kappa_\gamma+\lambda_\gamma) \ ,  \\
    Q_W^e &=& (-e/m_W^2)(\kappa_\gamma-\lambda_\gamma) \ ,  \\
    d_W   &=& (e/2m_W) 
              (\tilde{\kappa}_\gamma+\tilde{\lambda}_\gamma) \ ,  \\
    Q_W^m &=& (-e/m_W^2)
              (\tilde{\kappa}_\gamma-\tilde{\lambda}_\gamma) \ . 
\end{eqnarray*} 
In the $\SM$ the couplings at tree level are given by 
$\kappa_V = 1$ ($\Delta\kappa_V$=$\kappa_V$-1=0), 
$\lambda_V$=${\tilde \kappa}_V$=${\tilde \lambda}_V$=0. 
Because of the similarity of the ${\cal CP}$--conserving and 
${\cal CP}$--violating terms in the Lagrangian, the kinematic behavior of 
these terms is similar and the limits on both sets of anomalous couplings 
will be approximately the same. Therefore 
${\cal CP}$-violating terms will 
not be discussed explicitly. 
Also, unless stated, it will be assumed that 
$\Delta\kappa_\gamma = \Delta\kappa_Z$ and 
$\lambda_\gamma = \lambda_Z$.

A direct measurement of the moments of the $W$ boson, and thus of the 
gauge boson self-interactions, is 
possible through the study of gauge boson pair production. 
The cross sections for di-boson production, however, are all extremely small. 
For example, the predicted cross section times branching ratio 
for $W$-pair production with 
$WW \rightarrow \ell\ell\nu\nu$ ($\ell = e,\mu$) is about 0.5~pb 
and large integrated luminosities would be needed for a significant
measurement of the gauge couplings. 
The $\SM$ process of $W$-pair production, however, is characterized 
by large cancellations between the $s$ and $t$ channel 
production processes. The contributions from the $t$ channel
diagrams by themselves would violate unitarity. This 
implies that if the couplings deviate even modestly from their 
$\SM$ values,  
the gauge cancellations are destroyed and a large increase of the cross 
section is observed. Moreover, the differential distributions will be 
modified giving rise to gauge bosons with a large transverse boost since 
the largest gauge cancellations occur for highly boosted bosons.

A $WWV$ interaction Lagrangian with constant anomalous couplings
would thus violate unitarity at high energies and therefore
the coupling parameters must be modified to include
form factors~\cite{Baur}, that is, 
$\Delta\kappa (\hat{s}) = \Delta\kappa/(1+\hat{s}/\Lambda^2)^2 $ and 
$      \lambda(\hat{s}) =      \lambda/(1+\hat{s}/\Lambda^{2})^{2}$, 
where $\hat{s}$ is the square of the center of mass energy of the
subprocess. 
$\Lambda$ is a unitarity preserving form factor scale 
and indicates the scale at which the $\SM$ predictions are probed. 
In the next subsections different types of gauge boson pair production
will be discussed.

\subsection{$W$ Pair Production }

\D0 has searched for $W$-boson pair production
$\pbarp \rightarrow WW + X \rightarrow \ell\ell^\prime\nu\nu^\prime$
$(\ell\ell^\prime = ee/e\mu/\mu\mu)$~\cite{d0_ww}. 
The standard selection criteria for $W$-events have an overall 
efficiency for $W$-pair production of $\approx$ 0.07 and with an 
integrated luminosity of ${\cal L} \approx$ 14 pb$^{-1}$ 
$0.47 \pm 0.07$ events are expected from $\SM$ processes. 
The most significant background to this process is \ttbar production. 
Because of the additional two $b$-jets in \ttbar events, this background 
can be eliminated in a straightforward way by a cut on the hadronic activity 
in the event. \D0 applies a cut on the $p_T$ of the $WW$-system, 
$E_T^{HAD}  \,=\, | - ( \vec{E}_T^{\ell_1} \, + \, 
                        \vec{E}_T^{\ell_2} \, + \, 
                        \etmisv )| $, 
which is required to be less than 40 GeV. This requirement rejects about 
75\% of the \ttbar background and has an efficiency of 95\% for the 
expected $WW$ signal. 
The searches in the 
$ee\nu\nu$, $e\mu\nu\nu$ and $\mu\mu\nu\nu$ channels yield one signal 
event with an anticipated background of 0.56 $\pm$ 0.13 events. 
An upper limit on the $W$-pair production cross section of 
$\sigma(WW) < 87$ pb$^{-1}$ has been set at 95\% CL. 

With larger integrated luminosities it is possible to measure the 
$W$-pair production cross section.  
Based on an integrated luminosity of 
${\cal L} = $ 108 pb$^{-1}$ CDF has done an analysis similar to the \D0 
analysis searching for $W$-pairs in the di-lepton channel 
using a jet veto, that is, events with jets with 
$E_T > 10$~GeV
are rejected. The selection yields 5 signal events on 
a background of 1.2 $\pm$ 0.3 events. 
The measured $W$-pair production cross section is 
$\sigma$(\pbarp $\rightarrow WW) = (10.2^{+6.3}_{-5.1} \pm 1.6)$~pb, 
where the $\SM$ predicts 
$\sigma_{SM}$(\pbarp $\rightarrow WW) = (9.5 \pm 1.0)$~pb. 
It should be pointed out that the smallness of the cross sections in 
itself is a beautiful demonstration of the gauge cancellations in 
the $\SM$. 

\begin{figure}[t]
    \epsfxsize = 8.0cm
    \centerline{\epsffile[100 150 600 650]{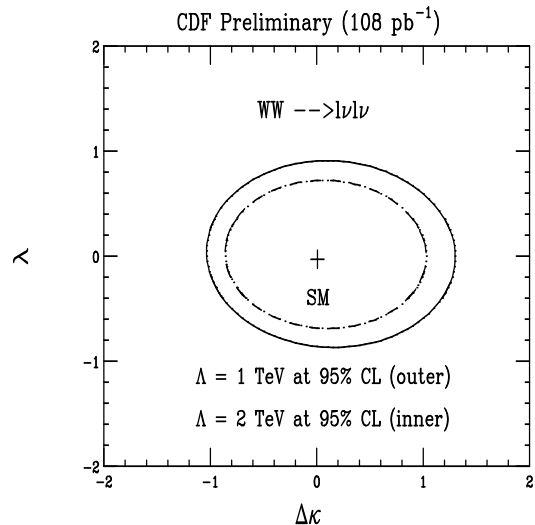}}
\caption{CDF exclusion contours in $\Delta\kappa$ and $\lambda$ obtained
         from the measurement of the $W$-pair production cross section 
         in the di-lepton channel for two different form factor scales, 
         assuming $\lambda_{\gamma} = \lambda_{Z}$ and
         $\Delta\kappa_\gamma = \Delta\kappa_{Z}$. }
\label{fig:cdf_ww}
\end{figure}

Since the cross section
increases very rapidly when the couplings deviate from their $\SM$ 
values, the measured 95\% CL upper limit on the cross section can be used 
to set limits on anomalous couplings. Figure~\ref{fig:cdf_ww} shows the 
CDF 95\% CL exclusion contours in $\Delta\kappa$ and $\lambda$ for two
different form factor scales, 
assuming 
$\lambda_{\gamma} = \lambda_{Z}$ and 
$\Delta\kappa_\gamma = \Delta\kappa_{Z}$. 
It is customary to quote limits on only one coupling, 
keeping the other couplings fixed to their $\SM$ value. 
These, so called, axis limits 
for a form factor scale of $\Lambda = 2$~TeV are 
$-1.0 < \Delta\kappa < 1.3~~~(\lambda = 0)$,   
$-0.9 < \lambda < 0.9~~~(\Delta\kappa = 0)$ 
for the CDF analysis, under the assumption that 
$\lambda_{\gamma} = \lambda_{Z}$ and 
$\Delta\kappa_\gamma = \Delta\kappa_{Z}$.

\subsection{$WW$ and $WZ$ Production }

Searches for particle production requiring two leptons in the 
final state always suffer in rate because of the small leptonic 
branching ratios. 
When in the analysis described in the previous subsection only one
lepton is required, a substantial increase in event rate is obtained
though at the cost of a much larger background. 
The background from $W/Z$+jet production to these processes 
is about 30 times higher than for the signal production. 
Given the distinct characteristics of anomalous couplings, this
background can be dealt with. 
Anomalous couplings modify the differential
distributions dramatically, especially the transverse momentum distribution 
of the $W$-boson. The ratio 
$\frac {\sigma_{WW}(p_T^W=200 \  {\rm GeV/c}) }
       {\sigma_{WW}(p_T^W=20 \ {\rm GeV/c})}$ 
is about $10^{-3}$, 
whereas for only modest deviations from $\SM$ couplings
($\Delta\kappa = 0,\lambda=1.0$) 
this ratio is about 0.5.  
By requiring the vector boson to have high transverse momentum 
the background is completely eliminated 
and a good sensitivity to anomalous couplings is retained. One 
completely loses sensitivity, however, to $\SM$ 
$WW/WZ$-production. 

Both CDF and \D0 have looked for $WW$ and $WZ$-production using
hadronic decay channels~\cite{cdf_ww_wz,d0_ww_wz}. The CDF analysis
proceeds by selecting events 
with one high $p_T$ lepton, large \etmis and 
2 jets with $E_T > 30$~GeV. 
Since the jets come from the hadronic decay of the gauge boson, their
invariant mass is required to be consistent with the gauge boson mass, 
$60 < m_{jj} < 110$ GeV/c$^2$. 
Since no distinction can be made between $WW$ and $WZ$-production in this 
selection, the sensitivity of the study was increased by including 
\pbarp$ \rightarrow WZ \rightarrow q{\overline q}^\prime\ell\ell$ events,
requiring the di-lepton invariant mass to reconstruct to the $Z$-boson mass. 
In the data sample, corresponding to a total integrated luminosity of 
110~pb$^{-1}$, no events are observed with 
$p_T^{jj} > 200$~GeV/c in the search region 
$60 < m_{jj} < 110$ GeV/c$^2$. 
A background of 0.8~events from $W/Z$+jet events
is expected and 0.1~events are predicted from $\SM$ processes. 
Limits on anomalous couplings can then be set based on the 
event rate yielding, for $\Lambda = 2$~TeV, 
\begin{center}
\begin{tabular}{ccc}
    $-0.5 <  \Delta\kappa < 0.6$   & $(\lambda = 0)$       & \\
    $-0.4 <  \lambda < 0.3$        & $(\Delta\kappa = 0)$  & .
\end{tabular}
\end{center}

%
%

The \D0 experiment has performed a similar analysis based on their
Run~1a data sample of 14~pb$^{-1}$, using only $W\rightarrow e\nu$
decays. The leptonic decays of the $Z$ are not considered in this
analysis. 
Since gauge bosons produced from anomalous self-interactions 
tend to have high $p_T$, the jets from such a high $p_T$ $W$ or $Z$
boson may not be well separated in space.
In order to maximize the detection efficiency of $W$ and $Z$ bosons with
high $p_T$, a small jet cone size of $\Delta R = 0.3$ was used in this
analysis. 
The detection efficiency for hadronic decays of $W$ and $Z$ bosons 
was estimated as a function of $p_T$ using Monte Carlo. 
The detection efficiency was found to be $\sim$60\%, 
approximately constant 
up to $p_T^{jj} = 350$ GeV/c. Differences in the estimated efficiencies 
from different 
Monte Carlo generators were included in the systematic uncertainty. 
The $p_T^{e\nu}$ spectrum of the final event sample of 84 events 
is of course dominated by background. 
The total number of background events was estimated to be $75.5\pm13.3$, 
with $12.2 \pm 2.6$ events coming from QCD multi-jet events and 
$62.2 \pm 13.0$ from W+jet events. The remaining small background
is mainly due to $\ttbar$ production. 
The SM prediction for $WW/WZ$ production was $3.2 \pm 0.6$ events. 

Because anomalous couplings not only affect the event rate but also
significantly alter differential distributions, 
better limits on anomalous couplings are obtained when
utilizing the full spectrum. 
\D0 has performed a maximum likelihood fit to the $p_T^{e\nu}$ spectrum
and, assuming equal $WWZ$ and $WW\gamma$ couplings, 
obtained the following limits at 95\%
confidence level: 
\begin{center}
\begin{tabular}{ccc}
    $-0.9 < \Delta\kappa < 1.1$   & $(\lambda = 0)$        & \\
    $-0.6 < \lambda < 0.7 $       & $(\Delta\kappa = 0)$   & , 
\end{tabular}
\end{center}
using $\Lambda = 1.5$~TeV. Comparing these limits to 
those obtained by CDF for the same process, but with five times the 
statistics using both electron and muon decays, 
shows the additional constraint that can obtained from 
the shape of the distribution. 

\begin{figure}[h]
\begin{center}
\begin{tabular}{cc}
    \epsfxsize = 4.0cm
    \epsffile{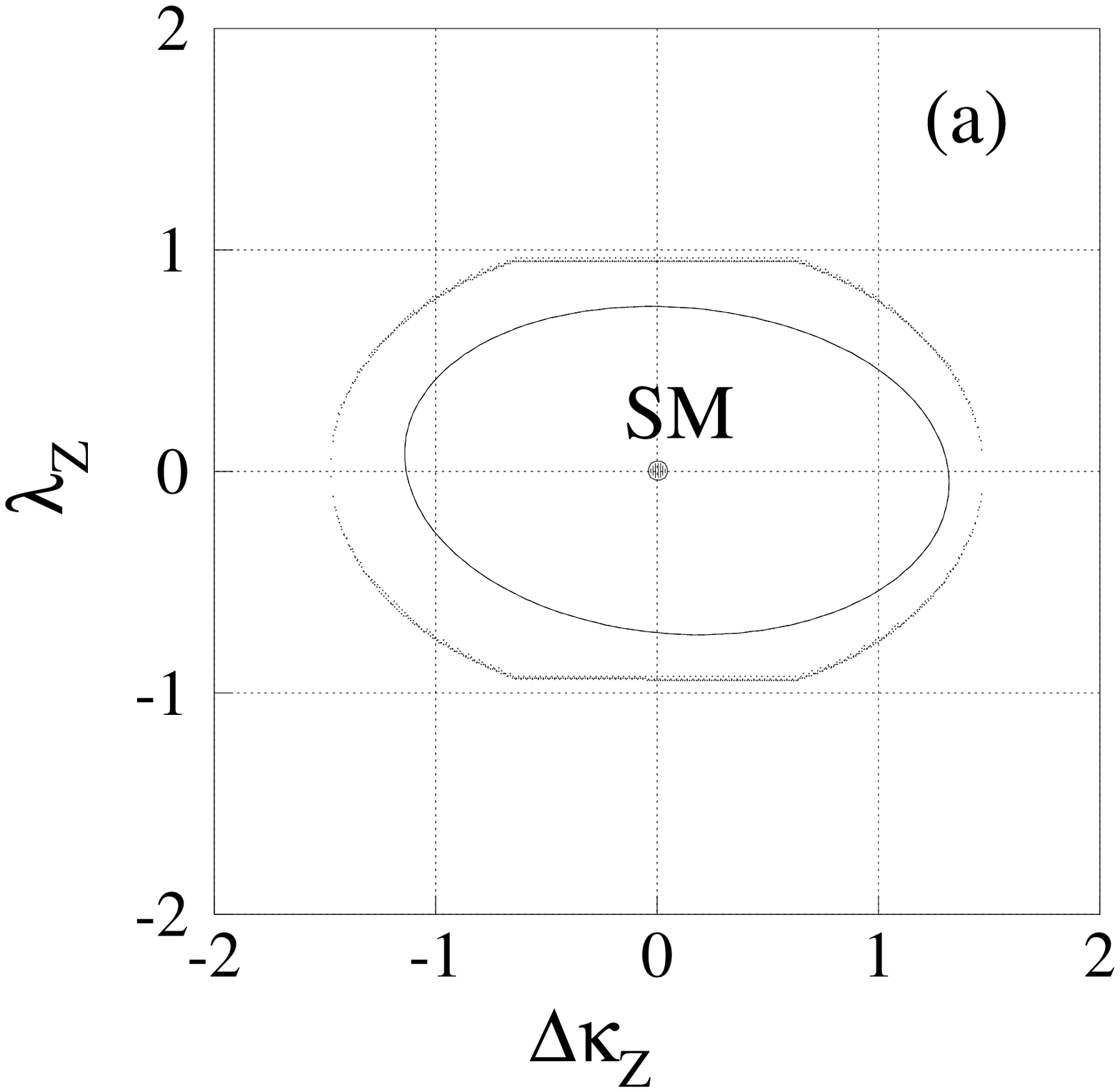}     &
    \epsfxsize = 4.0cm
    \epsffile{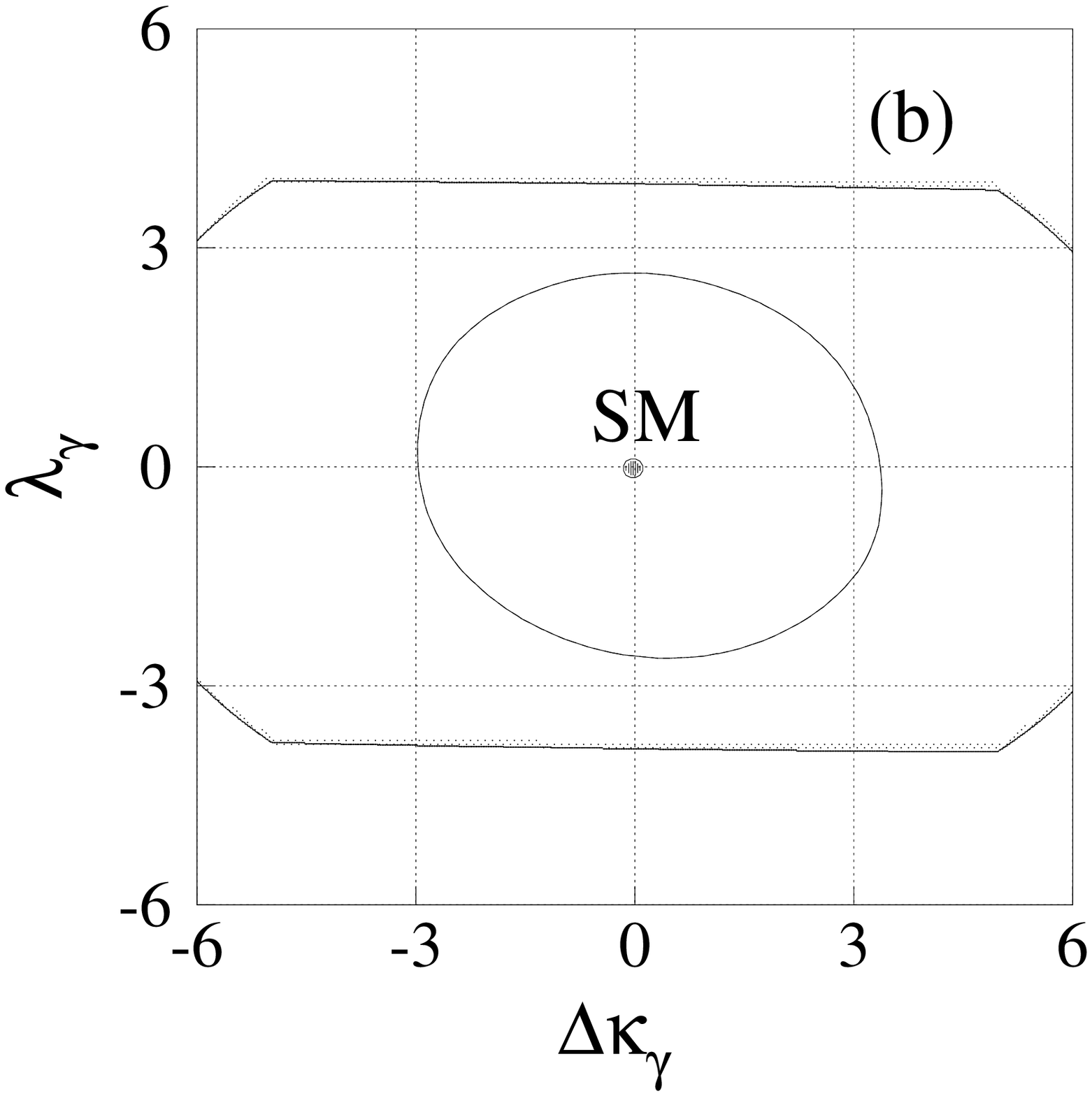}
\end{tabular}
\end{center}
\caption{Contour limits on anomalous coupling parameters at the 95\% CL
         (inner curves) and unitarity contours (outer curves) for \D0 
         assuming $\Lambda=1.5$~TeV for the process 
         $WW/WZ\rightarrow e\nu jj$. $\SM$ couplings have been assumed
         for 
         (a) $WW\gamma$ and 
         (b) $WWZ$ vertex. }
\label{fig:d0_ww_wz} 
\end{figure}

Since this analysis probes both $WW\gamma$ and $WWZ$ couplings, 
information can be obtained on the $WWZ$ coupling alone by setting 
the $WW\gamma$ couplings to their $\SM$ values. 
Fig.~\ref{fig:d0_ww_wz}a shows the contour limits when $\SM$ $WW\gamma$
couplings are assumed, whereas the $WWZ$ coupling was set to its
$\SM$ value in Fig.~\ref{fig:d0_ww_wz}b. 
The contours indicate that the analysis is more sensitive to the $WWZ$
coupling than the $WW\gamma$ coupling as expected from the larger
coupling strength of the $WWZ$ vertex. Also noteworthy is the
observation that the data confirms the existence of the $WWZ$ vertex. 

\subsection{$W\gamma$ Production }

The study of the production of photons in association with a $W$ also 
permits a study of the $WW\gamma$-vertex
\cite{ua2_wgamma,cdf_wgamma,d0_wgamma}. 
Most photons produced in association with a $W$, however, are radiated
off the initial or final state fermion. The only channel that allows for
a direct probe of the triple gauge boson vertex is the 
$s$-channel contribution of a photon radiated from a $W$. 
In the analyses $W\gamma$ events are selected by requiring, 
in addition to the regular 
$W$ selection criteria, an isolated photon with transverse energy 
$E_T^\gamma > 10\ (7)$ GeV for \D0 (CDF). Photons are detected in the 
pseudo-rapidity range 
$|\eta_\gamma | < 1.1$ for CDF and 
$|\eta_\gamma | < 1.1$ or $1.5 < |\eta_\gamma | < 2.5$ for \D0. 
The photon identification efficiencies are approximately 80\% for CDF 
and 75\% (58\%) for \D0 for the central (end) region.   
To reduce the contribution from radiative events the photon is 
required to be well separated from the lepton from the $W$-decay,
$\Delta R(\ell\gamma) > 0.7\,$.

\begin{table*}[t]
\begin{center}
\begin{tabular}{||l|cc|cc||} \hline\hline
    & \multicolumn{2}{c|} { \D0 }
    & \multicolumn{2}{c||}{ CDF }  \\ 
    & \multicolumn{2}{c|} { 87 pb$^{-1}$ }
    & \multicolumn{2}{c||}{ 67 pb$^{-1}$ }  \\ \hline
    & $W\gamma\rightarrow e\nu\gamma $  
    & $W\gamma\rightarrow \mu\nu\gamma $
    & $W\gamma\rightarrow e\nu\gamma $  
    & $W\gamma\rightarrow \mu\nu\gamma $
\\ \hline
$N_{\rm data}$  & 57                      & 70 
                & 75                      & 34                  \\
$N_{\rm bkg}$   & 15.2 $\pm$ 2.5          & 27.4 $\pm$ 4.7 
                & 16.1 $\pm$ 2.4          & 10.3 $\pm$ 1.2      \\
$N_{\rm sig}$   & $41.8^{+8.8}_{-7.5}$    & $42.6^{+9.7}_{-8.3}$  
                & $58.9 \pm 9.0 \pm 2.6 $ & $23.7 \pm 5.9 \pm 1.1 $     \\
$N_{\rm SM}$    & 43.6 $\pm$ 3.1          & 38.2 $\pm$ 2.8 
                & $53.5 \pm 6.8$          & $21.8 \pm 4.3  $     \\
\hline\hline 
\end{tabular}
\end{center}
\caption{Number of $W\gamma$ events observed in the data, expected 
         background and signal events. Also listed is the number of
         expected events for $\SM$ couplings. } 
\label{tab:wgamma} 
\end{table*}

\begin{figure}[h]
    \epsfxsize = 8.0cm
    \centerline{\epsffile{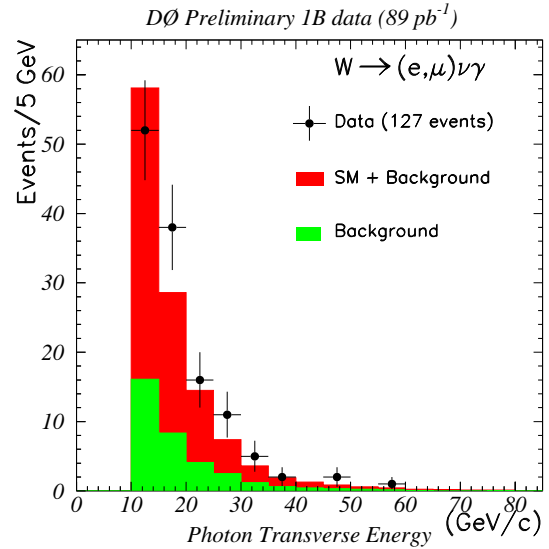}} 
\caption{$p_T^\gamma$ distribution of \D0 $W\gamma$ candidate events. } 
\label{fig:w_gamma_spectra}
\end{figure}

\begin{figure}[h]
    \epsfxsize = 8.0cm
    \centerline{\epsffile{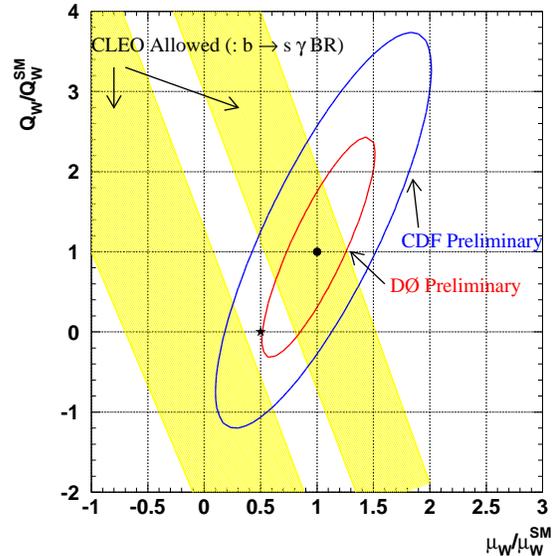}}
\caption{Limits on anomalous magnetic dipole and electric quadrupole 
         moments for the $W$ boson from CDF, \D0 and CLEO. }
\label{fig:wgamma_lim}
\end{figure}

The number of signal events, after background subtraction, and the number 
of expected events from $\SM$ processes are listed 
in table~\ref{tab:wgamma} for the electron and muon channels separately. 
Figure~\ref{fig:w_gamma_spectra} shows the distribution of the 
photon $p_T$-spectrum for \D0, together with the $\SM$ expectation. 
Good agreement with the prediction is observed and limits could be set
based on the event rate. 
As seen in the previous section, if the event statistics allows it, 
better limits on anomalous couplings are obtained by performing a 
maximum likelihood fit to a differential distribution. 
For $W\gamma$ production a binned maximum likelihood 
fit is performed to the 
$E_T^\gamma$-spectrum as function of the coupling constants. The 
last data bin is explicitly taken to be a zero-event bin. 
The limits thus obtained for a form-factor scale
$\Lambda=1.5$~TeV are 
\begin{center}
\begin{tabular}{ccc}
    $-1.0 < \Delta\kappa < 1.0$  & $(\lambda = 0)$       & (\D0) \\
    $-1.8 < \Delta\kappa < 2.0$  & $(\lambda = 0)$       & (CDF) \\
    $-0.3 < \lambda < 0.3$       & $(\Delta\kappa = 0)$  & (\D0) \\
    $-0.7 < \lambda < 0.6$       & $(\Delta\kappa = 0)$  & (CDF). 
\end{tabular}
\end{center}
The corresponding contours in magnetic dipole and electric quadrupole
moment, in units of the $\SM$ prediction for the moments, 
are shown in Fig.~\ref{fig:wgamma_lim}. 
A vanishing magnetic dipole moment and electric 
quadrupole moment of the $W$, corresponding to 
$\kappa = -\frac{1}{2}$ and 
$\lambda = -\frac{1}{2}$ is excluded at 99\% CL. 

The decay rate for $b\rightarrow s\gamma$ can also be used to set limits
on anomalous couplings since the process is sensitive to photon radiation off
the $W$-boson in the penguin diagram. The branching ratio has been measured
by CLEO to be 
$B(b\rightarrow s\gamma) = (2.32 \pm 0.57 \pm 0.35)\, 10^{-4}$ 
\cite{cleo_bsgamma}. The upper limit on this branching ratio excludes the 
outer regions in Fig.~\ref{fig:wgamma_lim}. 
The narrow region between the two
allowed CLEO bands is excluded by the lower limit. 

\subsection{Combined Result on $WW\gamma$ Coupling }

The studies of $W\gamma$ and $WW/WZ$ production are both
sensitive to the same $WW\gamma$ coupling. The analyses can thus be
combined to improve on the limits on anomalous couplings. 
When combining results, the correlation between the different 
analyses needs to be addressed. 
Some of the dominant common systematic uncertainties are 
due to the method of estimating the background and 
the uncertainty in structure functions and photon identification. 
The \D0 experiment has carried out a combined fit to the three
data sets corresponding to the 
$WW$, $WW/WZ$ and $W\gamma$ analyses from Run~1a. 
The significantly improved limits are: 
\begin{center}
\begin{tabular}{cc}
    $-0.7 < \Delta\kappa < -0.9$    & $(\lambda = 0)$         \\
    $-0.4 < \lambda < 0.4$          & $(\Delta\kappa = 0)$,
\end{tabular}
\end{center}
where it was assumed that the $WWZ$ couplings and the $WW\gamma$
couplings were equal. 
Note that this combined result is more stringent than the result from 
the \D0 $W\gamma$ analysis using the complete Run~1 data sample, 
showing the reach when all Tevatron results are combined. 

%
%

\subsection{$Z\gamma$ Production }

The $ZZ\gamma$ and $Z\gamma\gamma$ trilinear gauge
boson couplings are described in a way analogous to the 
$WWV$ couplings. 
These couplings, absent in the $\SM$, are suggested by some theoretical
models which imply new  physics~\cite{zg_review}. 
The most general Lorentz and gauge invariant $ZV\gamma$ vertex is 
described by eight coupling  parameters,  $h^V_i,~(i=1...4)$, 
where $V = Z,\gamma$~\cite{Hagiwara}. 
Combinations of the ${\cal CP}$--conserving (${\cal CP}$--violating) 
parameters $h^V_3$ and $h^V_4$ ($h^V_1$  and $h^V_2$)  
correspond  to the  electric  (magnetic) dipole and
magnetic (electric) quadrupole transition moments of the $ZV\gamma$ vertex.
Partial  wave  unitarity  of the   general $f\bar f \to  Z\gamma$ process
restricts the $ZV\gamma$ couplings uniquely to their vanishing 
$\SM$ values at
asymptotically  high energies~\cite{unitarity}.  Therefore,  the coupling
parameters  have to be  modified by  form-factors 
$h^V_i =  h^V_{i0} / (1 + \hat{s}/\Lambda^2)^n$, 
where  $\hat{s}$ is the square of the invariant mass
of the $Z\gamma$ system and $\Lambda$ is the form-factor scale. 
The energy dependence of the form factor is assumed to be 
$n = 3$ for $h^V_{1,3}$  and $n = 4$ for $h^V_{2,4}$~\cite{Baur_Z}. 
Such a choice yields the same asymptotic energy behavior for 
all the couplings.  

\begin{figure}[h]
    \epsfxsize = 8.cm
    \centerline{\epsffile{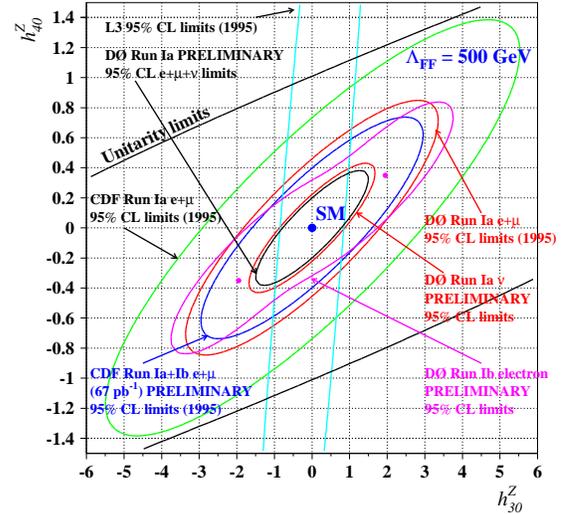}}
\caption{Limits on anomalous ${\cal CP}$-conserving $ZZ\gamma$ couplings from
         $Z(\ell\ell)\gamma$ and $Z(\nu\nu)\gamma$ production. 
         The dashed line is the unitarity contour for a form-factor scale 
         $\Lambda = 500$~GeV.}
\label{fig:zg_limits}
\end{figure}

The study of anomalous couplings in the process 
$Z\gamma\rightarrow \ell\ell\gamma$ follows the same lines as the 
$W\gamma$ analysis \cite{cdf_zgamma,d0_zgamma}. 
Table~\ref{tab:zgamma} lists the expected and
observed number of signal events for both experiments. 
The total cross section is seen to be in good agreement with the 
$\SM$ prediction. 
The sensitivity to anomalous couplings 
lies in the high $p_T^\gamma$ region. 
Three events with $p_T^\gamma > 60$~GeV/c are observed, one by CDF and
two by \D0. For \D0, the probability to observe at least two events with
$p_T^\gamma > 60$~GeV/c, given a total of 14 events observed, 
is 8.2\% and the events are consistent with a 
signal or background fluctuation within two standard deviations. 
Because of these high $p_T$ events, however, 
small non-vanishing anomalous couplings are preferred in the \D0 analysis. 
Their resulting exclusion contour from the Run~1b electron data 
is therefore slightly distorted~(see Fig.~\ref{fig:zg_limits}). 
Preliminary limits on 
anomalous couplings for a scale factor $\Lambda=500$~GeV 
from the di-electron analysis by \D0 and the di-lepton analysis by CDF 
are, at 95\% CL, 
\begin{center}
\begin{tabular}{ccc} 
    $-1.8 < h_{30}^Z < 1.8$ & $(h_{40}^Z = 0)$    & (\D0)     \\
    $-1.6 < h_{30}^Z < 1.6$ & $(h_{40}^Z = 0)$    & (CDF)     \\
    $-0.4 < h_{40}^Z < 0.4$ & $(h_{30}^Z = 0)$    & (\D0)     \\
    $-0.4 < h_{40}^Z < 0.4$ & $(h_{30}^Z = 0)$    & (CDF)     \\
\end{tabular}
\end{center}

\begin{table}[t]
\begin{center}
\begin{tabular}{|l|cc|cc|} 
\hline\hline
                & \multicolumn{2}{c|}{ \D0  }
                & \multicolumn{2}{c|}{ CDF  }  \\ 
                & \multicolumn{2}{c|}{ 89 pb$^{-1}$ }
                & \multicolumn{2}{c|}{ 67 pb$^{-1}$ } \\ \hline
                & $e$        & 
                & $e$        & $\mu$   \\ \hline
$N_{\rm data}$  & 14                        & 
                & 18                        & 13                        \\
$N_{\rm bkg}$   & 1.6 $\pm$ 0.5             & 
                & 0.9 $\pm$ 0.3             & 0.5 $\pm$ 0.1             \\
$N_{\rm Sig}$   & $12.4^{+4.8}_{-3.7} \pm 0.5$  & 
                & $17.1 \pm 5.7 $           & $12.5 \pm 3.6 $           \\
$N_{\rm SM }$   & $12.0 \pm 1.2$            & 
                & $16.2 \pm 1.8$            & $8.7  \pm 0.7$            \\
\hline 
\end{tabular}
\end{center}
\caption{Number of $Z\gamma$ events observed in the data, expected 
         background and signal events. Also listed is the number of 
         expected events for $\SM$ couplings. } 
\label{tab:zgamma} 
\end{table}

The \D0 experiment has recently performed a new analysis looking for the
decay $Z\gamma\rightarrow \nu\nu\gamma$. 
This channel has previously been studied only in 
$e^+e^-$-collisions~\cite{l3_zgamma}. 
Sensitivity to anomalous couplings in this channel is much  
higher than in the di-lepton decay modes due
to the higher decay rate into neutrinos and the absence  
of radiative $Z$ decay background. 
The overall background, however, is still extremely high, leading 
to very stringent event selection criteria. 
To reduce the background from $W$+jet events 
with the electron or jet being misidentified as a photon 
the $E_T^\gamma$ and \etmis were required to exceed 40~GeV. 
In addition, events with at least one jet with 
$E_T^j > 15$~GeV were rejected. 
The  remaining  background  was dominated  by cosmic rays and muons from
beam halo which radiated in the calorimeter. 
This background  was suppressed  by rejecting events with a 
reconstructed muon or a minimum ionizing trace 
in the  calorimeter close to the photon cluster.  
The residual background, which had roughly equal contributions from 
$W \to e\nu$ decays and muon bremsstrahlung, was derived from data. 

                                                                        
Four candidate events are observed on an expected background of  
$6.4 \pm 1.1$  events and a $\SM$ prediction of $1.8 \pm 0.2$ events. 
Although the signal-to-background ratio is less than one, 
the sensitivity to anomalous couplings is still high,
since the background is concentrated at low $E_T^\gamma$ while
the anomalous coupling contribution is almost flat in $E_T^\gamma$ up to
the kinematic threshold of the reaction.  
Limits on anomalous couplings were  set at 95\%  CL by a fit to 
the  $E_T^\gamma$ spectrum and gives 
$|h^Z_{30}| < 0.9$,  $|h^Z_{40}| < 0.2$. 
This represents a factor of two improvement compared to the combined 
\D0 Run~1a limits from the di-lepton analysis, 
based on the same luminosity~\cite{d0_zgamma}. 
A summary of all the limits is shown in 
Fig.~\ref{fig:zg_limits}~\cite{cdf_zgamma,d0_zgamma,l3_zgamma}. 
The L3 contour has a different orientation because of the different 
subprocess center of mass energy at which the events are produced.

\section{Conclusions}
A wide variety of properties of the $W$ and $Z$-bosons are now being 
studied at hadron colliders with ever increasing precision, at the 
highest energy scales achievable. 
All results, including the results from $\ee$ 
colliders~\cite{dpf96,swartz}, are in good agreement with the 
$\SM$. 
It is widely anticipated, though, that the $\SM$ is just an 
approximate theory and should eventually be replaced by a more 
complete and fundamental description of the underlying forces in nature. 
With the new data from LEP~2, SLD and the Tevatron, 
and with the planned upgrades of the accelerators as well as the
experiments, the projected uncertainties~\cite{snowmass_ewk} 
on some fundamental parameters should provide the 
tools to take another ever more critical look at the $\SM$, 
without any theoretical prejudice.

\section{Acknowledgements}
I would like to thank Debbie Errede, Paul Grannis, Young-Kee Kim 
and Darien Wood who have been very helpful.


\begin{thebibliography}{99}
\bibitem{old_masses} Although these mass values have been updated 
                     recently~\cite{dpf96}, these are the averages 
                     used for the published results.
\bibitem{angle} 
        The angle $\vartheta$ is the polar angle of the outgoing fermion 
        with respect to the incoming proton beam.
        Pseudorapidity is defined as $\eta = -\ln\tan{\vartheta\over 2}$. 
        An asterisk indicates quantities in the parton center of mass
        frame.
\end{thebibliography}
\end{document}